\documentclass[10pt,a4paper]{article}
\usepackage[utf8]{inputenc}
\usepackage[english]{babel}

\usepackage{amsmath}
\usepackage{amsfonts}
\usepackage{amssymb}

\usepackage[hidelinks]{hyperref}

\usepackage[left=2cm,right=2cm,top=2cm,bottom=2cm]{geometry}

\usepackage{authblk}

\usepackage{feynmp}
\newcommand{\pd}{\partial}
\newcommand{\abs}[1]{\left\lvert #1 \right\rvert}
\newcommand{\ce}{\varepsilon}

\title{On two body gravitational scattering within perturbative gravity}
\author[1,2]{\href{https://orcid.org/0000-0001-7099-0861}{B. Latosh} \thanks{ \href{mailto:latosh.boris@ibs.re.kr}{latosh.boris@ibs.re.kr} } }
\affil[1]{Particle Theory  and Cosmology Group, Center for Theoretical Physics of the Universe, Institute for Basic Science (IBS), Daejeon, 34126, Korea}
\affil[2]{Bogoliubov Laboratory of Theoretical Physics, JINR, Dubna 141980, Russia}
\author[3]{A. Yachmenev}
\affil[3]{Dubna State University, Universitetskaya str. 19, Dubna 141982, Russia}
\date{CTPU-PTC-23-12}

\begin{document}

\maketitle

\begin{abstract}
  We study the gravitational scattering of two scalar particles up to the one-loop order. We calculate off-shell amplitudes and operators with the recently developed package FeynGrav. We obtained the tree-level amplitude and found it consistent with classical results in relativistic and non-relativistic cases. We obtained explicit expressions for the vertex operator and the scattering amplitude at the one-loop level off-shell. They are consistent with the previous results obtained in the low energy limit. Analysis of the scattering amplitudes and operators is given. We outline the further development achievable with the FeynGrav and other computational packages.
\end{abstract}

\section{Introduction}

The standard perturbative approach to quantum gravity produces a non-renormalisable theory. At the one-loop level, pure general relativity without matter is ultraviolet finite on-shell \cite{tHooft:1974toh}. The theory generates new higher derivative operators $R^2$, $R_{\mu\nu}^2$, and $R_{\mu\nu\alpha\beta}^2$. Because on-shell graviton states satisfy vacuum Einstein equations $R_{\mu\nu} =0$, divergent contributions proportional to $R^2$ and $R_{\mu\nu}^2$ vanish in on-shell amplitudes. The Riemann tensor squared $R_{\mu\nu\alpha\beta}^2$ on-shell is reduced to the Gauss-Bonnet invariant, which is a complete derivative in $d=4$, so the corresponding divergent contribution also vanishes on-shell. This feature is lost for off-shell amplitudes and higher loop corrections. Similarly, if matter or higher curvature terms are added to the theory, even on-shell one-loop amplitudes receive ultraviolet divergent contributions \cite{Deser:1974cz,Deser:1974cy,Goroff:1985sz,Latosh:2020jyq}.

Perturbative quantum gravity shall be considered an effective theory. It can be renormalised in a way that differs from the renormalisation of quantum electrodynamics or quantum chromodynamics. Within both quantum electrodynamics and quantum chromodynamics, it is possible to define a finite set of operators with which all ultraviolet divergences at all levels of perturbation theory are subtracted. The same thing is impossible within perturbative quantum gravity because, at each level of perturbation theory, it generates new higher dimensional operators. Although extending the theory of gravity to the ultraviolet region is impossible, it is always possible to subtract ultraviolet divergences from a given scattering amplitude and derive significant results. Therefore, studying this theory and obtaining valuable information about its quantum features is possible.

The main aim of this paper is to study the gravitational scattering of two scalar particles of different masses. This scattering takes a rather unique place within perturbative quantum gravity. First, the corresponding amplitudes are relatively easy to calculate, even off-shell. Second, the result can be directly compared with the classical scattering. Third, the process provides a concrete example of quantum gravity calculations that can be studied in great detail. Let us elaborate on these statements. 

The gravitational scattering of scalar particles takes a special place within quantum gravity for a few reasons. First and foremost, the structure of scalar-tensor gravity is well-understood. The most general scalar-tensor theory admitting second-order field equations is given by the Horndeski theory \cite{Horndeski:1974wa,Kobayashi:2011nu}. It constrains the variety of healthy non-minimal couplings between a scalar field and gravity and reduces the number of operators that can present in a microscopic gravity action. The theory contains operators influencing the speed of gravitational waves in the contemporary Universe \cite{Kase:2018aps}. The magnitude of the corresponding operators, including those generated at the loop level, is now constrained by the LIGO data \cite{Latosh:2020jyq,LIGOScientific:2017zic,Ezquiaga:2017ekz}.

Secondly, scalar particle scattering amplitudes contain fewer terms, as they do not involve Lorentz or spinor indices. The number of terms in a scattering amplitude increases with the number of particles involved and the loop order. Spinor, vector, or graviton in external states provides additional complication because the resulting amplitude shall involve the corresponding indices and respect the corresponding symmetries. Partially, this issue can be negated with the usage of on-shell amplitudes, but they are not free from disadvantages, which we discuss in the following paragraph. Moreover, quantum field theory is not reduced to the on-shell amplitudes alone, so calculations of off-shell amplitudes are necessary. Therefore, analysing amplitudes that only involve scalar external states is the most direct subject to investigate.

Lastly, techniques for off-shell amplitudes calculations are significantly less developed than those for on-shell amplitudes. There are numerous techniques used to calculate on-shell scattering amplitude for massless particles. Even their brief review lies far beyond the scope of this paper \cite{Dixon:1996wi,Elvang:2013cua,Vanhove:2021zel,Travaglini:2022uwo}. Many of these tools cannot be used for off-shell amplitudes or amplitudes involving massive particles. For instance, the spinor helicity formalism, which plays an enormous role in scattering amplitude calculations, is well-understood and widely used only for on-shell amplitudes. It is well-developed and widely used for massless particles, while its generalisation for massive particles was only developed relatively recently \cite{Arkani-Hamed:2017jhn}. Consequently, calculations of off-shell amplitudes involving massive particles heavily rely on the standard Feynman rules. Recently developed computational package FeynGrav \cite{Latosh:2022ydd,Latosh:2023zsi}, which is based on widely used FeynCalc \cite{Shtabovenko:2016sxi,Shtabovenko:2020gxv,Shtabovenko:2016whf,Patel:2015tea}, provides a tool to operate with Feynman rules for gravity.

The discussed gravitational scattering amplitudes can be directly compared with classical cases. In the non-relativistic case, one can compare the tree-level cross section calculated within perturbative quantum gravity with a generalisation of the Rutherford formula. In the relativistic case, one can construct a meaningful comparison with a scattering of a scalar probe particle on a Schwarzschild black hole. Within general relativity, it is possible to construct a description of the gravitational scattering of two massive scalar particles. The description is given by the Einstein-Infeld-Hoffmann equations \cite{Einstein:1938yz,Einstein:1940mt,Straumann:2013spu}. These equations contain non-potential interactions because relativistic field equations contain momenta of particles. To avoid introducing non-potential interactions, we will only consider the scattering of a massive test particle on a Schwarzschild black hole. This problem is similar to scattering on a rigidly fixed body, and the mass hierarchy strongly suppresses all non-potential interactions. We numerically recover the corresponding differential cross-section and study its behaviour at small scattering angles. These results provide a meaningful comparison between the tree-level results obtained in perturbative quantum gravity and classical results.

With the power of the contemporary computational tools mentioned above, it is possible to analyse the discussed scattering in detail. We will operate mainly with an amplitude of a gravitational scattering of two scalars of different masses and the vertex operator describing the interaction of a scalar with gravity. In both cases, we study the structure of ultraviolet and infrared divergences with the help of \mbox{Package-X} \cite{Patel:2015tea,Patel:2016fam}.

We will also study the low-energy limit of the theory. By the low-energy limit, we mean the limit when all external momentum of an amplitude approaches zero. Let us note that the discussed limit shall not be confused with the widely known study of quantum gravity infrared features (although they are strongly related). The infrared features of quantum gravity lie far beyond the scope of this paper, as they are related to its fundamental features \cite{Donoghue:1994dn,Donoghue:2001qc,BjerrumBohr:2002kt,Bjerrum-Bohr:2002fji,Bjerrum-Bohr:2013bxa,Bjerrum-Bohr:2014zsa,Bjerrum-Bohr:2016hpa,Bjerrum-Bohr:2018xdl,Bjerrum-Bohr:2021vuf}. This paper focuses on more practical aspects of quantum gravity and uses that limit only to obtain a Newtonian-like approximation of the theory to compare it with the well-known results. Because of this, by the low-energy limit, we mean only the leading order contribution of a given scattering amplitude when all external momenta approach zero, and we shall not discuss its fundamental features further.

According to the BPHZ theorem \cite{Bogoliubov:1957gp,Hepp:1966eg,Zimmermann:1969jj}, the ultraviolet divergent part of amplitude can only contain terms that are analytic functions of external momenta. Such a structure of an amplitude allows one to extract the low-energy limit of a theory. Since the divergent part of a given amplitude can only be an analytic function of external momenta, it either vanishes or remains constant in the low-energy limit. In turn, the ultraviolet finite part contains non-analytic functions of external momenta, which are dominant in the low-energy limit. In particular, such non-analytic terms are responsible for power-law corrections to the Newton potential, which empirical data can verify \cite{Battista:2014oba,Battista:2014ija,Battista:2015zta,Battista:2015wwa,Battista:2017xlm,Tartaglia:2018bjc,Battista:2020qqp}.

This paper also implements an approach based on scattering cross-sections. Scattering cross sections are directly observable, free from ambiguities typical for gauge theories, and are Lorentz invariant. We compare the differential cross-sections of gravitational scattering within classical and quantum theories to establish a correspondence between them. Firstly, we will calculate the differential cross-section for the classical non-relativistic case and obtain an analogue of the Rutherford formula for the gravitational cross-section for massive scalar particles. Secondly, we address the classical relativistic case by considering the gravitational scattering of a massive test particle on a Schwarzschild black hole. We numerically recover the corresponding differential cross-section and study its behaviour at small scattering angles. Finally, we will study the gravitational scattering of two massive scalar particles within perturbative quantum gravity. We show that the differential cross section evaluated within perturbative quantum gravity at the tree level correctly recovers the Rutherford formula in the suitable limit and accounts for relativistic corrections to small angle scattering.

The paper is organised as follows. In Section \ref{Section_non-relativistic_scattering}, we briefly discuss the classical non-relativistic gravitational scattering and obtain a generalisation of the Rutherford formula. In Section \ref{Section_relativistic_scattering}, we discuss relativistic gravitational scattering. We show that a scattering of a test particle of a Schwarzschild black hole allows one to account for relativistic corrections for the potential scattering. It is possible to obtain an analytical formula expression of the scattering angle $\theta$ as a function of the impact parameter $\rho$ and initial particle momentum $p$. However, resolving it for $\rho = \rho(p,\theta)$ is impossible. We use numerical calculations to recover the differential cross-section for small angles. Section \ref{Section_perturbative_quantum_gravity} explicitly calculates the gravitational scattering differential cross-section for two scalar particles of different masses. We show that it admits the Rutherford limit and correctly recovers the small scattering angle data obtained for the relativistic scattering. In Section \ref{Secition_one-loop}, we discuss one-loop scattering amplitude and one-loop vertex operator and analyse them. In Section \ref{Section_conclusion}, we discuss the obtained results and outline further development achievable with FeynGrav.

\section{Rutherford scattering}\label{Section_non-relativistic_scattering}

We start the discussion with the classical non-relativistic case. It is similar to the well-known Rutherford scattering, discussed in many textbooks \cite{Landau1976Mechanics,Arnold1997mathematical,goldstein2002classical}, so we will not discuss it in great detail. This section provides an analogue of the Rutherford formula for non-relativistic $2\to 2$ gravitational scattering.

The well-known Lagrangian describes the classical non-relativistic motion of two gravitationally attracted particles:
\begin{align}
  \mathcal{L} = \cfrac{p_1^2}{2\,m_1} + \cfrac{p_2^2}{2\, m_2} + \cfrac{G\,m_1\,m_2}{\abs{\vec{r}_1-\vec{r}_2}}\, .
\end{align}
Here $G$ is the Newton constant; $m_1$ and $m_2$ are masses of particles; $\vec{p}_1$, $\vec{p}_2$ and $\vec{r}_1$, $\vec{r}_2$ are momenta and radius vectors of particles. The centre of mass motion is factorised via a suitable choice of variables:
\begin{align}
  \mathcal{L} = \cfrac{P^2}{2\,M} + \cfrac{p^2}{2\,\mu} + \cfrac{G\,m_1\,m_2}{r}\,.
\end{align}
Here $M = m_1 +m_2$ is the total mass, $\mu= (m_1^{-1} + m_2^{-1})^{-1}$ is the reduced mass, and momenta and radius vectors are related as follows:
\begin{align}
  \begin{cases}
    \vec{r} = \vec{r}_1 - \vec{r}_2 ,\\
    \vec{R} = \cfrac{m_1\, \vec{r}_1 + m_2 \, \vec{r}_2}{m_1+m_2}\,,
  \end{cases}
  & &
  \begin{cases}
    \vec{P} = \vec{p}_1+\vec{p}_{\,2} ,\\
    \vec{p} = \cfrac{m_2}{m_1+m_2}~ \vec{p}_1 - \cfrac{m_1}{m_1+m_2} ~ \vec{p}_2 \,.
  \end{cases}
\end{align}
The problem is reduced to an effective one-body problem. A suitable equation of motion is obtained from the energy conservation law:
\begin{align}\label{energy_conservation_1}
  E = \cfrac{\mu}{2} \, \big( \dot{r} \big)^2 + \cfrac{L^2}{2\,\mu\,r^2} - \cfrac{G\,m_1\,m_2}{r} \,.
\end{align}
Here, $E$ is the centre of mass energy, and $L$ is the centre of mass angular momentum. The conservation law produces the following master equation:
\begin{align}\label{master_1}
  \left(\cfrac{d\frac{1}{r}}{d\varphi}\right)^2 = \cfrac{2\,\mu\,E}{L^2} + \left(\cfrac{G\,\mu\,m_1\,m_2}{L^2}\right)^2 - \left(\cfrac{1}{r}  - \cfrac{G\,\mu\,m_1\,m_2}{L^2}\right)^2.
\end{align}
Non-relativistic master equation \eqref{master_1} admits an analytical solution:
\begin{align}\label{master_1_differentials}
  d\varphi = d \arcsin \cfrac{\frac{1}{r}-\frac{G\,\mu\,m_1\,m_2}{L^2}}{\sqrt{ \frac{2\,\mu\,E}{L^2} + \left(\frac{G\,\mu\,m_1\,m_2}{L^2}\right)^2 }}\,.
\end{align}

To solve a scattering problem is to relate the scattering angle $\theta$ with the impact parameter $\rho$. This relation can be extracted from \eqref{master_1_differentials} via a direct integration. To define the integration limits, one shall consider motion from the spacial infinity to the turnaround radius. At the turnaround radius, the radial velocity of a particle vanishes. Because of this, the value of the turnaround radius is found from \eqref{energy_conservation_1}:
\begin{align}
  r_\text{turnaround} = \cfrac{G\,m_1\,m_2}{2\,E} \left[-1+\sqrt{1+\cfrac{L^2}{2\,\mu \,E}\left(\cfrac{2\,E}{G\,m_1\,m_2}\right)^2}\,\,\right].
\end{align}
The angle changes from zero to the turnaround angle $\varphi_0$ which, in turn, is related to the scattering angle $\theta$:
\begin{align}
  \theta + 2\,\varphi_0 = 2\pi\,.
\end{align}
With such integration limits equation \eqref{master_1_differentials} reads:
\begin{align}
  \varphi_0 = \cfrac{\pi}{2} + \arcsin \cfrac{\frac{G\,\mu\,m_1\,m_2}{L^2}}{\sqrt{ \frac{2\,\mu\,E}{L^2} + \left(\frac{G\,\mu\,m_1\,m_2}{L^2}\right)^2 }} \,.
\end{align}
The angular momentum $L$ and the energy $E$ shall be expressed in terms of the impact parameter $\rho$ and the centre of mass momentum $p_\text{cm}$ as follows:
\begin{align}
  E &= \cfrac{p_\text{cm}^2}{2\,\mu} \,, & L &= \rho \, p_\text{cm}\,.
\end{align}
These formulae provide the desirable relation:
\begin{align}
  \rho^2 = \left( \cfrac{G\,\mu \, m_1 \, m_2}{p_\text{cm}^2}\right)^2 \, \cot^2 \cfrac{\theta}{2}\,.
\end{align}
The following analogue of the Rutherford formula gives the differential cross section:
\begin{align}\label{differential_cross_section_NR}
  \cfrac{d \sigma}{d \Omega} = \cfrac{1}{4} \, \left(G\,\mu\, m_1\, m_2\right)^2 \cfrac{1}{p_\text{cm}^4\,\sin^4\frac{\theta}{2}} \,.
\end{align}
The expression \eqref{differential_cross_section_NR} is the main result of this section. The following sections show that it appears as the suitable limit in relativistic and quantum cases.

The expression \eqref{differential_cross_section_NR} has several important features. First, expression \eqref{differential_cross_section_NR} is given in the centre of the mass frame using the centre of mass momentum $p_\text{cm}$ and the centre of mass scattering angle $\theta$. Therefore, it can only be compared to cross-sections obtained in the centre of the mass frame. Secondly, the expression vanishes in the massless limit ($m_1\to 0$ or $m_2\to 0$ ). Consequently, relativistic cross-sections describing the gravitational scattering of massless particles will not reproduce \eqref{differential_cross_section_NR}.  On the contrary, cross sections describing the gravitational scattering of massive particles shall recover \eqref{differential_cross_section_NR} as the leading contribution in $p_\text{cm} \sim 0$ limit. The same logic holds for the quantum case. One shall expect to recover \eqref{differential_cross_section_NR} within perturbative gravity for the gravitational scattering of two scalar particles of non-vanishing masses.

\section{Black hole scattering}\label{Section_relativistic_scattering}

Let us turn to a discussion of relativistic gravitational scattering. A direct approach to this problem would require a Lagrangian that describes the gravitational interaction between two relativistic particles, which can be obtained from the post-Newtonian limit of general relativity. This will lead to a complicated set of equations that cannot be solved analytically for a scattering problem. The Einstein-Infeld-Hoffmann equations \cite{Einstein:1938yz,Einstein:1940mt,Straumann:2013spu}, derived from such a Lagrangian, contain non-potential terms that describe interactions dependent on particles' momenta, adding to the complexity of the problem.

Alternatively, one can study the scattering of a test particle on a black hole, treating it as a two-body problem with one body fixed in place, neglecting any back reaction of the black hole on the test particle. Birkhoff's theorem justifies this simplification for small angle cross section. Small angle scattering occurs when two particles interact at a significant distance, far greater than the typical Schwarzschild radius. In that case, the theorem guarantees that the Schwartzschild metric describes the external gravitational field of a point-like particle. Indeed, this makes the analogy fail when particles are scattered for large angles or participate in a non-elastic scattering. However, these are not the cases of consideration. The main aim of this section is to extract the leading contribution to the gravitational scattering alone to prove that all the calculations admit the same limit.

This simplification allows for the derivation of an analytical expression for the scattering angle and corresponding differential cross section by considering only potential interactions, omitting terms of order $\mathcal{O}\left(m/M\right)$ and $\mathcal{O}\left(p/M\right)$, where $m$ is the test particle mass, $p$ is the test particle three-momentum, $M$ is the black hole mass.

Let us proceed with the proposed approach and state the scattering problem. We use the Schwarzschild black hole as the rigidly fixed body and consider the scattering of a generic massive scalar particle. Within such a setup, both bodies have non-vanishing masses and have no spin, which makes a complete analogy with the previously considered non-relativistic case. 

For the sake of generality, we start with a generic metric of a stationary non-rotating black hole:
\begin{align}
  ds^2 = A(r)\, c^2 \, dt^2 - B(r) \, dr^2 - D(r) \,\left[d\theta^2 +\sin^2\theta \, d\varphi\right] .
\end{align}
The corresponding equations of motion for a massive particle are derived in the standard way \cite{Chandrasekhar:1985kt,Perlick:2021aok}. They produce the following relativistic generalisation of the master equation \eqref{master_1}:
\begin{align}\label{master_2}
  \left(\cfrac{d\frac{1}{r}}{d\varphi}\right)^2 = \cfrac{1}{r^4} \,\cfrac{D}{B} \,\left[ mc^2 \, \cfrac{m\,D}{L^2}\left\{\,\left(\cfrac{E}{mc^2}\right)^2\cfrac{1}{A}-1\right\} -1 \right] .
\end{align}
Here $m$ is the test particle mass, $E$ is the test particle energy, and $L$ is the test particle angular momentum. For the Schwarzschild black hole metric functions are defined as follows:
\begin{align}
  A &= (B)^{-1} = 1-\cfrac{2 \, G \, M}{c^2\, r} \,, & D(r) &= r^2 ,
\end{align}
with $M$ being the black hole mass. Consequently, equation \eqref{master_2} is reduced to the following form:
\begin{align}
  \left(\cfrac{d\frac{1}{r}}{d\varphi}\right)^2 &=\cfrac{\cfrac{1}{c^2}\,\left(E^2 - (m c^2)^2\right)}{L^2} + \left(\cfrac{G\,M \, m^2}{L^2}\right)^2 - \left(\cfrac{1}{r}  - \cfrac{G\,M\,m^2}{L^2}\right)^2 + \cfrac{2\,G\,M}{c^2} \, \cfrac{1}{r^3} \,.
\end{align}
This equation shall be simplified even further. Firstly, one shall use the dispersion relation $E^2 = p^2 c^2 +m^2 c^4$ to express the first term via the impact parameter $\rho$ alone:
\begin{align}
  \cfrac{\cfrac{1}{c^2}\,\left(E^2 - (m c^2)^2\right)}{L^2} = \cfrac{p^2}{L^2} = \cfrac{1}{\rho^2}\,.
\end{align}
Secondly, one should note that the reduced mass of the system matches the mass of the test particle (assuming $m \ll M$):
\begin{align}
  \cfrac{1}{\mu} = \cfrac{1}{M}+ \cfrac{1}{m} \overset{M \gg m}{\sim} \cfrac{1}{m}\,.
\end{align}
These considerations lead to the following relativistic generalisation of the master equation \eqref{master_1}:
\begin{align}\label{master_2_GR}
  \left(\cfrac{d\frac{1}{r}}{d\varphi}\right)^2 &=\cfrac{1}{\rho^2} + \left(\cfrac{G\,\mu\,M\,m}{L^2}\right)^2 - \left(\cfrac{1}{r}  - \cfrac{G\,\mu\,M\,m }{L^2}\right)^2 + \cfrac{2\,G\,M}{c^2} \, \cfrac{1}{r^3} \,.
\end{align}
The first three terms of this equation match terms of equation \eqref{master_1} up to a change of notations. The last term describes the leading order relativistic correction to Newton's potential. 

Solution of \eqref{master_2_GR} is obtained via elliptic integrals \cite{Collins:1973xf,Synge:1960ueh}.
Let us introduce the following notations:
\begin{align}
  L =& p \, \rho \, , & r_0 =& \cfrac{2 \, G\, M}{c^2} \,, & u =& \cfrac{r_0}{r} \, , &  b =& \cfrac{r_0}{\rho} \,, & \ce =& \cfrac{p }{mc}\, .
\end{align}
Here $b$ describes the inverse impact parameter measured in the horizon radius units, and parameter $\ce$ serves as a measure of the relativistic nature of the system. Namely, if $\ce \geq 1$, then the particle momentum $p \geq mc$ and the particle is relativistic. In this notations equation \eqref{master_2_GR} is reduced to the following form:
\begin{align}\label{master_2_u}
  \left(\cfrac{d u}{d\varphi} \right)^2 = u^3 - u^2 + \cfrac{b^2}{\ce^2}\, u + b^2 .
\end{align}
To begin, we will cover the process of integrating this equation and then discuss the limits of integration.

To integrate \eqref{master_2_u}, one shall factorise the right-hand side of this equation through its roots:
\begin{align}
  u^3 - u^2 + \cfrac{b^2}{\ce^2}\, u + b^2 = (u-u_1) (u-u_2)(u-u_3)\,.
\end{align}
All roots shall be treated as a function of $b$ and $\ce$. We are only interested in the case when all roots are real and $u_1\leq 0 \leq u_2\leq u_3$. If the equation has only one real root, the particle is captured by the black hole, and this case lies beyond the scope of this discussion. This imposes the following constraint on parameters $b$ and $\ce$:
\begin{align}
  0\leq \ce \leq \infty \,,  & &  0 \leq b \leq \cfrac{1}{2\sqrt{2}}\,\sqrt{\ce^2 - 18\,\ce^4-27\ce^6 + \ce^2 \sqrt{1+\ce^2} (1+9\ce^2)^{3/2}} \leq \cfrac{2}{3\sqrt{3}} \,.
\end{align}

To proceed with the integration, \eqref{master_2_u} should be brought to the following form:
\begin{align}
  \left(\cfrac{d\sqrt{\frac{u-u_1}{u_2-u_1}}}{d\,\frac12\,\varphi\, \sqrt{u_3-u_1}}\right)^2 = \left[1-\cfrac{u-u_1}{u_2-u_1}\right] \, \left[1- \cfrac{u_2-u_1}{u_3-u_1} \, \cfrac{u-u_1}{u_2-u_1}\right].
\end{align}
In this form, variables can be split, and the integration is performed:
\begin{align}\label{master_2_u_integral}
  \int ~d\,\frac12\varphi \sqrt{u_3-u_1} = \int \cfrac{ d\sqrt{\frac{u-u_1}{u_2-u_1}} }{\sqrt{ \left[1-\frac{u-u_1}{u_2-u_1}\right] \, \left[1- \frac{u_2-u_1}{u_3-u_1} \, \frac{u-u_1}{u_2-u_1}\right]  } }\,.
\end{align}
The right-hand side of this equation is reduced to the elliptic integral $F$:
\begin{align}
  F(\phi,m) \overset{\text{def}}{=} \int\limits_0^\phi \cfrac{d\theta}{\sqrt{1-m \sin^2 \theta}} =\int\limits_0^{\sin\phi} \cfrac{dt}{\sqrt{(1-t^2) (1- m\, t^2)}}\,.
\end{align}

In full analogy with the previous case, it is helpful to study the motion of a test particle from spacial infinity up to the turnaround radius. The initial position of the particle $r\to\infty$ corresponds to $u=0$. The position of the turnaround radius is defined by roots of the right-hand side of \eqref{master_2_u}. Root $u_2$ defined the turnaround radius because it is the smallest root in the physical domain $u>0$. Consequently, the left-hand side of \eqref{master_2_u_integral} shall be integrated from $\varphi=0$ to the turnaround angle $\varphi_0$. The right-hand side of \eqref{master_2_u_integral} shall be integrated from $u=0$ to $u=u_2$. This provides the following analytic expression for the turnaround angle:
\begin{align}
  \varphi_0 = \cfrac{2}{\sqrt{u_3-u_1}} \left[ F\left(\frac{\pi}{2} , \frac{u_2-u_1}{u_3-u_1}\right) - F\left(\arcsin \sqrt{\frac{-u_1}{u_3-u_1}} , \frac{u_2-u_1}{u_3-u_1}\right) \right] .
\end{align}
Consequently, an analytic expression for the scattering angle $\theta$ is also found:
\begin{align}\label{scattering_angle_relativistic_case}
  \theta (b,\varepsilon) = \abs{\pi - \cfrac{4}{\sqrt{u_3-u_1}} \left[ F\left(\frac{\pi}{2} , \frac{u_2-u_1}{u_3-u_1}\right) - F\left(\arcsin \sqrt{\frac{-u_1}{u_3-u_1}} , \frac{u_2-u_1}{u_3-u_1}\right) \right]  } .
\end{align}
Again, let us highlight that parameters $b=r_0/\rho$ and $\varepsilon = p/(mc)$ enter this expression implicitly via roots $u_1$, $u_2$, and $u_3$. An explicit analytic expression for roots can be obtained via Cardano's formula.

Equation \eqref{scattering_angle_relativistic_case} provides an analytic expression relating $\theta$ with $b$ and $\ce$. To the best of our knowledge, it is impossible to solve \eqref{scattering_angle_relativistic_case} analytically for $b = b(\theta,\ce)$. We use numerical methods to obtain the desirable dependence $b=b(\theta,\ce)$. If function $b=b(\theta,\ce)$ is known, the differential cross section shall be calculated as follows:
\begin{align}
  \cfrac{d\sigma}{d\Omega} = \cfrac{\rho}{\sin\theta} \, \abs{\cfrac{d\rho}{d\theta}} = \cfrac{r_0^2}{\sin\theta}\,\cfrac{1}{b^3} \abs{\cfrac{d b}{d\theta} } \,.
\end{align}
Therefore, it is more convenient to operate with the reduced cross-section
\begin{align}
  \tilde\sigma &= \cfrac{\sigma}{r_0^2} \, , &  \cfrac{d\tilde\sigma}{d\Omega}  &= \cfrac{1}{\sin\theta} \, \cfrac{1}{b^3} \, \abs{\cfrac{db}{d\theta}}\,.
\end{align}

Numerical calculations show that the reduced differential cross section has the following leading order behaviour in the small angle limit:
\begin{align}\label{BH_cross_section_small_angle}
  \cfrac{d \tilde{\sigma} }{d\Omega} = \cfrac{1}{\theta^4} \, \left(\cfrac{1}{\ce^2} + 2\right)^2 + \mathcal{O}\left(\theta^{-3}\right).
\end{align}
Let us highlight the following features of the obtained expressions. Firstly, \eqref{BH_cross_section_small_angle}  has a finite massless limit $\ce\to\infty$, proving the result's self-consistency. Secondly, \eqref{BH_cross_section_small_angle} can be directly compared with the non-relativistic case. The expression \eqref{BH_cross_section_small_angle} was obtained in the laboratory system, but within the used setup ($M\gg m$), the centre of mass momentum $p_\text{cm}$ match the laboratory frame momentum $p$:
\begin{align}
\cfrac{p^2}{2m} =  \cfrac{p_\text{cm}^2}{2\mu} \simeq \cfrac{p_\text{cm}^2}{2 m} +\mathcal{O}\left(\cfrac{m}{M}\right) .
\end{align}
This allows one to bring \eqref{differential_cross_section_NR} to the laboratory frame:
\begin{align}
  \cfrac{d\sigma}{d\Omega} &=\cfrac14\,\left(G\,\mu\,m_1\,m_2\right)^2 \cfrac{1}{p_\text{cm}^4\,\sin^4\frac{\theta}{2} } \\
  &= \cfrac14\,\left(\cfrac{G\,M^2\,m^2}{p_\text{cm}^2}\right)^2 \cfrac{1}{\sin^4\frac{\theta}{2}} = \cfrac14\, \left(\cfrac{2\, G\, M}{c^2}\right)^2 \cfrac{1}{\left(\frac{p_\text{cm}}{m c}\right)^4} \cfrac{1}{\sin^4\frac{\theta}{2}} \simeq \left(\cfrac{2\, G\, M}{c^2}\right)^2 \, \cfrac{1}{\theta^4}\,\left(\cfrac{1}{\ce^2}\right)^2+\mathcal{O}\left(\theta^{-3}\right)\,. \nonumber
\end{align}
This result matches the non-relativistic limit ($\ce \to 0$) of \eqref{BH_cross_section_small_angle}. Therefore, the approach used in this section is self-consistent and consistent with the non-relativistic limit.

In summary, there is currently no known analytical method for deriving the impact parameter as a function of the scattering angle. However, the numerical value of the differential cross-section can always be calculated for any given impact parameter and momentum. Additionally, formula \eqref{BH_cross_section_small_angle} provides information about small angle scattering, which is consistent with the non-relativistic results \eqref{differential_cross_section_NR} derived in the previous section.

\section{Tree-level scattering}\label{Section_perturbative_quantum_gravity}

Finally, let us discuss the case of perturbative quantum gravity. Perturbative quantum gravity is a quantum theory of small metric perturbations $h_{\mu\nu}$, associated with gravitons, propagating about the flat background $\eta_{\mu\nu}$. The complete spacetime metric $g_{\mu\nu}$ accounts both for the background and for the perturbations:
\begin{align}\label{the_perturbative_expansion}
  g_{\mu\nu} = \eta_{\mu\nu} + \kappa \, h_{\mu\nu}\,.
\end{align}
Here $\kappa = \sqrt{32 \, \pi \, G_N}$ is the gravitational coupling expressed via the Newton constant $G_N$. It shall be noted that this procedure can be generalised for the case of a non-flat background as well \cite{Christensen:1979iy}. The flat background is more suitable for the studied case of particle scattering. The formula \eqref{the_perturbative_expansion} is a finite expression and not an infinite expansion. Still, it spawns infinite expansions for $g^{\mu\nu}$, $\sqrt{-g}$, Christoffel symbols, and Riemann tensor. Consequently, the microscopic action of general relativity (or any other gravity model) presents an infinite perturbative expansion containing infinite interaction terms.

Explicit expressions for such interaction rules are required for perturbative calculations. In \cite{Latosh:2022ydd}, an algorithm constructing explicit expressions for such Feynman rules was found (see also \cite{Prinz:2020nru,Sannan:1986tz,DeWitt:1967uc}). The algorithm is implemented in the FeynGrav package \cite{Latosh:2022ydd,Latosh:2023zsi} that extends the functionality of "FeynCalc" \cite{Shtabovenko:2016sxi,Shtabovenko:2020gxv,Mertig:1990an}. All further calculations will be performed with FeynGrav. A detailed discussion of the package and its usage is presented in \cite{Latosh:2022ydd} and will not be discussed further.

Gravitational scattering of two scalar particles of different masses provides an analogy with the cases discussed in previous sections. In the non-relativistic case \eqref{differential_cross_section_NR}, both particles explicitly have spin $0$ and different non-vanishing masses. In the relativistic case \eqref{BH_cross_section_small_angle}, the particle has spin $0$ and an arbitrary mass, which can also be set to zero. The black hole has zero angular momentum, so it can be conceived as an analogy of a particle with spin $0$. Its mass is not completely arbitrary, as it must be much bigger than the mass of the test particle. Because of this, the gravitational scattering of two scalar particles of different masses serves as a suitable generalisation of the previous cases. This section shows that the corresponding matrix exactly reproduces results \eqref{differential_cross_section_NR} and \eqref{BH_cross_section_small_angle}.

It is more convenient to perform calculations in the centre of the mass frame. In this frame, four momenta of the particles are defined as follows:
\begin{align}
  \begin{cases}
    p_1 = \left(\sqrt{m_1^2+p_\text{cm}^2} ,0,0,p_\text{cm}\right),\\
    p_2 = \left(\sqrt{m_2^2+p_\text{cm}^2} ,0,0,-p_\text{cm}\right),\\
    p_3 = \left(\sqrt{m_1^2+p_\text{cm}^2} , p_\text{cm}\,\sin\theta ,0, p_\text{cm}\,\cos\theta \right) ,\\
    p_4 = \left(\sqrt{m_2^2+p_\text{cm}^2} , -p_\text{cm}\, \sin\theta ,0, -p_\text{cm}\,\cos\theta \right) .
  \end{cases}
\end{align}
Here, $p_1$ and $p_3$ are momenta of the first particle with mass $m_1$ before and after the scattering; $p_2$ and $p_4$ are momenta of the second particle with mass $m_2$ before and after the scattering; $p_\text{cm}$ is the centre of mass three-momentum; $\theta$ is the centre of mass scattering angle. The following expressions give Mandelstam variables:
\begin{align}
  \begin{cases}
    s = (p_1+p_2)^2   &=\left(\sqrt{m_1^2+p_\text{cm}^2}+\sqrt{m_2^2+p_\text{cm}^2}\right)^2 \,, \\
    t = \,(p_1-p_3)^2 &= - 4 \, p_\text{cm}^2 \, \sin^2 \frac{\theta}{2} \,, \\
    u = (p_1-p_4)^2   &=\left(\sqrt{m_1^2+p_\text{cm}^2}-\sqrt{m_2^2+p_\text{cm}^2}\right)^2 - 4 \, p_\text{cm}^2 \, \cos^2 \frac{\theta}{2} \,.
  \end{cases}
\end{align}

The matrix scattering element in the leading order is given by the following expression:
\begin{align}
  \begin{gathered}
    \begin{fmffile}{Diags/D01}
      \begin{fmfgraph}(40,40)
        \fmfleft{L1,L2}
        \fmfright{R1,R2}
        \fmf{dashes}{L1,VL,L2}
        \fmf{dashes}{R1,VR,R2}
        \fmf{dbl_wiggly}{VL,VR}
        \fmfdot{VL,VR}
      \end{fmfgraph}
    \end{fmffile}
  \end{gathered}
  = \mathcal{M}= -i\,\cfrac{\kappa^2}{4\, t} \Bigg( s(s+t) - (2\,s+t)(m_1^2+m_2^2) + m_1^4 + m_2^4 \Bigg)\,.
\end{align}
The corresponding differential cross section is given by the following expression where all the standard kinematic factors are taken into account \cite{Bilenky:1982tw,Weinberg:1995mt,Peskin:1995ev}:
\begin{align}
  \begin{split}\label{cross_section_with_factors}
    \cfrac{d\sigma}{d\Omega} = &\cfrac{\sqrt{m_1^2+p_\text{cm}^2} \, \sqrt{m_2^2 + p_\text{cm}^2}}{ \sqrt{  \left(p_\text{cm}^2 + \sqrt{m_1^2+p_\text{cm}^2} \, \sqrt{m_2^2 + p_\text{cm}^2} \right)^2- m_1^2\,m_2^2}} \\
    & \times \cfrac{p_\text{cm}}{64 \,\pi^2\, \sqrt{m_1^2+p_\text{cm}^2} \, \sqrt{m_2^2 + p_\text{cm}^2} \left(\sqrt{m_1^2+p_\text{cm}^2} + \sqrt{m_2^2 + p_\text{cm}^2}\right)} \abs{\mathcal{M}}^2 
  \end{split}\\
  \begin{split}\label{Quantum_cross_section_tree_level}
    = &\cfrac{G^2}{4} \,\cfrac{1}{\sin^4\frac{\theta}{2}}\,\, \cfrac{\left(m_1^2 \,m_2^2\, + 2\, p_\text{cm}^2\, (m_1^2+m_2^2) + 4\, p_\text{cm}^2\, \cos^2\frac{\theta}{2}\left(p_\text{cm}^2 + \sqrt{m_1^2+p_\text{cm}^2} \, \sqrt{m_2^2 + p_\text{cm}^2} \right) \right)^2}{p_\text{cm}^3\, \left(\sqrt{m_1^2+p_\text{cm}^2} + \sqrt{m_2^2+p_\text{cm}^2}\right)\, \sqrt{  \left(p_\text{cm}^2 + \sqrt{m_1^2+p_\text{cm}^2} \, \sqrt{m_2^2 + p_\text{cm}^2} \right)^2- m_1^2\,m_2^2}}~.
  \end{split}
\end{align}
In \eqref{cross_section_with_factors}, the first factor is due to the normalisation of the unit incoming particle flow; the second factor is introduced for the sake of the Lorentz invariance.

The formula \eqref{Quantum_cross_section_tree_level} describes the tree-level cross-section of two scalar particles of different masses and constitutes one of the main results of this paper. However, one shall verify that \eqref{Quantum_cross_section_tree_level} admits the suitable limits obtained in previous Sections. To compare \eqref{Quantum_cross_section_tree_level} with \eqref{differential_cross_section_NR} one should take the leading contribution of \eqref{Quantum_cross_section_tree_level} in $p_\text{cm}\to 0$ limit:
\begin{align}
  \begin{split}
    \cfrac{d\sigma}{d\Omega} = &\cfrac14\, \left( G \, \mu\,m_1\,m_2 \right)^2 \,\cfrac{1}{p_\text{cm}^4 \sin^4\frac{\theta}{4}} \Bigg[ 1 + p_\text{cm}^2 \, \left(\cfrac{4}{m_1^2} + \cfrac{4 \cos\theta +3}{m_1\,m_2} + \cfrac{4}{m_2^2}  \right)\\
      &+ p_\text{cm}^4\,\left( \cfrac{4}{m_1^4}+\cfrac{5 \,(8 \cos\theta+5)}{4\, m_1^3\, m_2}+\cfrac{8 \cos^2 \theta + 16 \cos\theta + 25}{2 \,m_1^2\, m_2^2}+\cfrac{5\, (8 \cos\theta+5)}{4\,m_1\, m_2^3}+\cfrac{4}{m_2^4}\right) + \mathcal{O}(p_\text{cm}^6) \Bigg]\,.
  \end{split}
\end{align}
The leading term of this expression matches \eqref{differential_cross_section_NR} exactly. To compare \eqref{Quantum_cross_section_tree_level} with \eqref{BH_cross_section_small_angle}, a different limit should be taken. Firstly, notations shall be modified to match conventions used in the previous section:
\begin{align}
  m_1 &\to m \,, & m_2 &\to M \,, & p_\text{cm} &\to \ce\, m \,.
\end{align}
Secondly, one should take the leading order in $\theta \to 0$ and $m/M \to 0$:
\begin{align}
  \begin{split}
    \cfrac{d\sigma}{d\Omega} =&\cfrac{(2\,G\,M)^2}{\theta^4} \left[  \left(2+\cfrac{1}{\ce^2}\right)^2 +  2\, \sqrt{1+\ce^2}\,\left(4-\cfrac{1}{\ce^4}\right)\,\cfrac{m}{M}  +\mathcal{O}\left(\left(\frac{m}{M}\right)^2\right)\right] \\
    &+ \cfrac{(2\,G\,M)^2}{\theta^2} \left[\cfrac16\, \left(2+\cfrac{1}{\ce^2}\right)^2 -\cfrac13\,\sqrt{1+\ce^2} \left(2+\cfrac{1}{\ce^2}\right)\left(4+\cfrac{1}{\ce^2}\right)\,\cfrac{m}{M} +\mathcal{O}\left(\left(\frac{m}{M}\right)^2\right)\right] \\
    & +(2\,G\,M)^2 \left[ \cfrac{11}{720}\,\left(2+\cfrac{1}{\ce^2}\right)^2 - \cfrac{1}{360} \,\sqrt{1+\ce^2}\,\left(76 + \cfrac{60}{\ce^2} + \cfrac{11}{\ce^4}\right) \, \cfrac{m}{M} +\mathcal{O}\left(\left(\frac{m}{M}\right)^2\right) \right] + \mathcal{O}(\theta^2)\,.
  \end{split}
\end{align}
The leading order contribution exactly matches \eqref{BH_cross_section_small_angle}. This proves that \eqref{Quantum_cross_section_tree_level} is consistent with results obtained in the previous sections.

\section{One-loop scattering}\label{Secition_one-loop}

We turn to the discussion of one-loop amplitudes. All the amplitude discussed below were calculated with FeynGrav \cite{Latosh:2022ydd,Latosh:2023zsi}. They contain many terms; for instance, the expression for the one-loop vertex operator contains $11356$ terms. The main reason for creating and developing packages is to handle many terms that are almost impossible to manage manually. Presenting the calculations in print is impossible due to the lengthy expressions involved. To make it possible to examine and use the explicit form of operators involved in the present research, we published them separately in open access \cite{latosh2023}. We use Package-X \cite{Patel:2015tea,Patel:2016fam}, which provides a tool to operate with explicit expressions for loop integrals, and package FeynHelpers \cite{Shtabovenko:2016whf} that provides an interface to use it within FeynCalc.

At the one-loop level, the following diagrams contribute to the gravitational scattering of two scalars of different masses
\begin{align}\label{Amplitude_one-loop}
  \begin{split}
    \begin{gathered}
      \begin{fmffile}{Diags/D0}
        \begin{fmfgraph}(30,30)
          \fmftop{T1,T2}
          \fmfbottom{B1,B2}
          \fmf{dashes}{B1,V1}
          \fmf{dashes}{B2,V2}
          \fmf{dashes}{T1,V1}
          \fmf{dashes}{T2,V2}
          \fmf{phantom,tension=2}{V1,V,V2}
          \fmfblob{20}{V}
        \end{fmfgraph}
      \end{fmffile}
    \end{gathered}
    \hspace{5pt} \overset{\text{def}}{=} \hspace{5pt}
    \begin{gathered}
      \begin{fmffile}{Diags/D1}
        \begin{fmfgraph}(30,30)
          \fmfbottom{B1,B2}
          \fmftop{T1,T2}
          \fmf{dashes}{B1,V1,T1}
          \fmf{dashes}{B2,V2,T2}
          \fmf{dbl_wiggly}{V1,V2}
          \fmfdot{V1}
          \fmfblob{10}{V2}
        \end{fmfgraph}
      \end{fmffile}
    \end{gathered}
    ~ + ~
    \begin{gathered}
      \begin{fmffile}{Diags/D2}
        \begin{fmfgraph}(30,30)
          \fmfbottom{B1,B2}
          \fmftop{T1,T2}
          \fmf{dashes}{B1,V1,T1}
          \fmf{dashes}{B2,V2,T2}
          \fmf{dbl_wiggly}{V1,V2}
          \fmfdot{V2}
          \fmfblob{10}{V1}
        \end{fmfgraph}
      \end{fmffile}
    \end{gathered}
    ~ + ~
    \begin{gathered}
      \begin{fmffile}{Diags/D3}
        \begin{fmfgraph}(30,30)
          \fmfbottom{B1,B2}
          \fmftop{T1,T2}
          \fmf{phantom,tension=0.5}{L1,R1,R2,L2,L1}
          \fmf{dashes}{B1,L1}
          \fmf{dashes}{B2,R1}
          \fmf{dashes}{T1,L2}
          \fmf{dashes}{T2,R2}
          \fmfdot{L1,L2,R1,R2}
          \fmffreeze
          \fmf{dashes}{L1,L2}
          \fmf{dashes}{R1,R2}
          \fmf{dbl_wiggly}{L1,R1}
          \fmf{dbl_wiggly}{L2,R2}
        \end{fmfgraph}
      \end{fmffile}
    \end{gathered}
    ~ + ~
    \begin{gathered}
      \begin{fmffile}{Diags/D4}
        \begin{fmfgraph}(30,30)
          \fmfbottom{B1,B2}
          \fmftop{T1,T2}
          \fmf{phantom,tension=0.5}{L1,R1,R2,L2,L1}
          \fmf{dashes}{B1,L1}
          \fmf{dashes}{B2,R1}
          \fmf{dashes}{T1,L2}
          \fmf{dashes}{T2,R2}
          \fmfdot{L1,L2,R1,R2}
          \fmffreeze
          \fmf{dashes}{L1,L2}
          \fmf{dashes}{R1,R2}
          \fmf{dbl_wiggly}{L1,R2}
          \fmf{dbl_wiggly}{L2,R1}
        \end{fmfgraph}
      \end{fmffile}
    \end{gathered}
    ~ + ~
    \begin{gathered}
      \begin{fmffile}{Diags/D5}
        \begin{fmfgraph}(30,30)
          \fmfbottom{B1,B2}
          \fmftop{T1,T2}
          \fmf{phantom,tension=.5}{V1,V0,V2}
          \fmf{phantom,tension=.2}{V1,V2}
          \fmf{dashes}{B1,V1}
          \fmf{dashes}{T1,V2}
          \fmf{dashes}{B2,V0,T2}
          \fmfdot{V0,V1,V2}
          \fmffreeze
          \fmf{dashes}{V1,V2}
          \fmf{dbl_wiggly}{V1,V0}
          \fmf{dbl_wiggly}{V2,V0}
        \end{fmfgraph}
      \end{fmffile}
    \end{gathered}
    ~ + ~
    \begin{gathered}
      \begin{fmffile}{Diags/D6}
        \begin{fmfgraph}(30,30)
          \fmfbottom{B1,B2}
          \fmftop{T1,T2}
          \fmf{phantom,tension=.5}{V1,V0,V2}
          \fmf{phantom,tension=.2}{V1,V2}
          \fmf{dashes}{B2,V1}
          \fmf{dashes}{T2,V2}
          \fmf{dashes}{B1,V0,T1}
          \fmfdot{V0,V1,V2}
          \fmffreeze
          \fmf{dashes}{V1,V2}
          \fmf{dbl_wiggly}{V1,V0}
          \fmf{dbl_wiggly}{V2,V0}
        \end{fmfgraph}
      \end{fmffile}
    \end{gathered}
    ~ + ~
    \begin{gathered}
      \begin{fmffile}{Diags/D7}
        \begin{fmfgraph}(30,30)
          \fmftop{T1,T2}
          \fmfbottom{B1,B2}
          \fmf{dashes}{B1,V1,T1}
          \fmf{dashes}{B2,V2,T2}
          \fmf{dbl_wiggly,right=1}{V1,V2,V1}
          \fmfdot{V1,V2}
        \end{fmfgraph}
      \end{fmffile}
    \end{gathered}
  \end{split}
\end{align}
We use the same kinematic as in the previous section for this scattering. The first two diagrams describe the contribution of the one-loop vertex operator given by the following:
\begin{align}\label{Vertex_function_one-loop}
  \begin{gathered}
    \begin{fmffile}{Diags/G}
      \begin{fmfgraph}(30,30)
        \fmftop{T1,T2}
        \fmfbottom{B}
        \fmf{dashes}{T1,V,T2}
        \fmf{dbl_wiggly}{B,V}
        \fmfv{decor.shape=circle,decor.filled=shaded,decor.size=15}{V}
      \end{fmfgraph}
    \end{fmffile}
  \end{gathered}
  \overset{\text{def}}{=}
  \begin{gathered}
    \begin{fmffile}{Diags/G1}
      \begin{fmfgraph}(30,30)
        \fmftop{T1,T2}
        \fmfbottom{B}
        \fmf{dbl_wiggly,tension=2}{B,VB}
        \fmf{dashes,tension=2}{T1,VT1}
        \fmf{dashes,tension=2}{T2,VT2}
        \fmf{phantom}{VB,VT2,VT1,VB}
        \fmfdot{VB,VT1,VT2}
        \fmffreeze
        \fmf{dashes}{VT1,VT2}
        \fmf{dbl_wiggly}{VT1,VB,VT2}
      \end{fmfgraph}
    \end{fmffile}
  \end{gathered}
  +
  \begin{gathered}
    \begin{fmffile}{Diags/G2}
      \begin{fmfgraph}(30,30)
        \fmftop{T1,T2}
        \fmfbottom{B}
        \fmf{dbl_wiggly,tension=2}{B,VB}
          \fmf{dashes,tension=2}{T1,VT1}
          \fmf{dashes,tension=2}{T2,VT2}
          \fmf{phantom}{VB,VT2,VT1,VB}
          \fmfdot{VB,VT1,VT2}
          \fmffreeze
          \fmf{dbl_wiggly}{VT1,VT2}
          \fmf{dashes}{VT1,VB,VT2}
      \end{fmfgraph}
    \end{fmffile}
  \end{gathered}
  +
  \begin{gathered}
    \begin{fmffile}{Diags/G3}
      \begin{fmfgraph}(30,30)
        \fmftop{T1,T2}
        \fmfbottom{B}
        \fmf{dashes,tension=2}{T1,VT,T2}
        \fmf{dbl_wiggly,left=1}{VT,VB,VT}
        \fmf{dbl_wiggly,tension=2}{B,VB}
        \fmfdot{VT,VB}
      \end{fmfgraph}
    \end{fmffile}
  \end{gathered}
  +
  \begin{gathered}
    \begin{fmffile}{Diags/G4}
      \begin{fmfgraph}(30,30)
        \fmftop{T1,T2}
        \fmfbottom{B}
        \fmf{dashes}{T1,V,T2}
        \fmf{dbl_wiggly,tension=2}{V,B}
        \fmffreeze
        \fmf{phantom}{V,VV,T1}
        \fmffreeze
        \fmf{dbl_wiggly,left=1}{V,VV}
        \fmfdot{V,VV}
      \end{fmfgraph}
    \end{fmffile}
  \end{gathered}
  +
  \begin{gathered}
    \begin{fmffile}{Diags/G5}
      \begin{fmfgraph}(30,30)
        \fmftop{T1,T2}
        \fmfbottom{B}
        \fmf{dashes}{T1,V,T2}
        \fmf{dbl_wiggly,tension=2}{V,B}
        \fmffreeze
        \fmf{phantom}{V,VV,T2}
        \fmffreeze
        \fmf{dbl_wiggly,right=1}{V,VV}
        \fmfdot{V,VV}
      \end{fmfgraph}
    \end{fmffile}
  \end{gathered}
  +
  \begin{gathered}
    \begin{fmffile}{Diags/G6}
      \begin{fmfgraph}(30,30)
        \fmftop{T1,T,T2}
        \fmfbottom{B}
        \fmf{dashes,right=.3}{T1,V,T2}
        \fmf{dbl_wiggly,tension=2}{V,B}
        \fmffreeze
        \fmf{phantom,tension=2}{T,VT}
        \fmf{phantom}{V,VT}
        \fmffreeze
        \fmf{dbl_wiggly,left=.7}{V,VT,V}
        \fmfdot{V}
      \end{fmfgraph}
    \end{fmffile}
  \end{gathered}
\end{align}

We address the different parts of the problem in the following subsections. Firstly, we discuss the one-loop vertex operators and their features. Secondly, we discuss the complete amplitude and contributions given by different sub-graphs. Lastly, we obtain the low-energy limit of the amplitude and find it consistent with the previous results.

\subsection{One-loop vertex operator}\label{Section_one-loop_vertex_operator}

We use the following notation for the one-loop vertex operator:
\begin{align}
  \nonumber \\
  i\, \Gamma_{\mu\nu} (p_1,p_2,k,m)= \hspace{20pt}
  \begin{gathered}
    \begin{fmffile}{Diags/G7}
      \begin{fmfgraph*}(40,40)
        \fmftop{T1,T2}
        \fmfbottom{B}
        \fmf{dashes}{T1,V,T2}
        \fmf{dbl_wiggly}{B,V}
        \fmfv{decor.shape=circle,decor.filled=shaded,decor.size=15}{V}
        \fmflabel{$\mu\nu,k$}{B}
        \fmflabel{$p_1$}{T1}
        \fmflabel{$p_2$}{T2}
      \end{fmfgraph*}
    \end{fmffile}
  \end{gathered}  \\ \nonumber
\end{align}
In the repository \cite{latosh2023} file ``\texttt{Graviton-Scalar\_One-Loop\_Vertex\_Operator.dat}'' contains the explicit expression in the FeynCalc format. All momenta are off-shell in that expression, and only momentum conservation is used to simplify it. Let us also note that the operator vanishes in massive and massless cases when all the momenta are placed on the shell.

The operator involves the following Passarino-Veltman integrals:
\begin{align}
  \begin{split}
    & A_0(m^2)\, , \, B_0(p_1^2,0,m^2) \,, \, B_0(p_2^2,0,m^2) \,, \, B_0(k^2,0,0)\,,\,B_0(k^2,m^2,m^2)\\
    & C_0 (p_1^2,p_2^2,k^2,m^2,0,m^2)\,,\,C_0 (p_1^2,p_2^2,k^2,0,m^2,0).
  \end{split}
\end{align}
Among these integrals, only $A_0$ and $B_0$ contain ultraviolet divergences. Infrared divergences appear in both massive and massless cases. In the massive case $m \not =0$, the operator receives the divergence in a single integral when the graviton becomes soft:
\begin{align}\label{Vertex_operator_IR_1}
  \lim\limits_{k\to 0} C_0 (p_1^2, p_2^2, k^2, 0, m^2\not=0,0) \bigg|_\text{IR} = \cfrac{1}{\varepsilon_\text{IR}} ~ \cfrac{1}{p_1^2 - p_2^2} ~ \ln \cfrac{1 - \frac{p_2^2}{m^2} }{ 1 - \frac{p_1^2}{m^2}} \,.
\end{align}
Here, $\varepsilon_\text{IR}$ is the infrared-divergent factor. In the massless case, the operator receives infrared divergences in the same integral if any particle is soft:
\begin{align}\label{Vertex_operator_IR_2}
  \lim\limits_{k\to 0} C_0 (p_1^2, p_2^2, k^2, 0, 0,0) \bigg|_\text{IR} = \cfrac{1}{\varepsilon_\text{IR}} ~ \cfrac{1}{p_1^2 - p_2^2} ~ \ln \cfrac{p_2^2}{p_1^2}  \,.
\end{align}
Similar relations hold for $p_1\to 0$ and $p_2 \to 0$ cases because the integral is symmetric with respect to momentum transmutation:
\begin{align}
  C_0 (a,b,c,0,0,0) =  C_0 (b,a,c,0,0,0) =  C_0 (a,c,b,0,0,0) =  C_0 (c,b,a,0,0,0) .
\end{align}
These infrared divergences are typical for the standard quantum field theory, and they can be removed by the standard methods \cite{Weinberg:1965nx}. 

Lastly, let us turn to a discussion of ultraviolet divergences. In the most general case, the operator takes the following form:
\begin{align}
  i\,\Gamma_{\mu\nu} = \alpha \,\eta_{\mu\nu} + \beta \left[ (p_1)_\mu (p_2)_\nu + (p_2)_\mu (p_1)_\nu \right] + \Delta_1 \, (p_1)_\mu (p_1)_\nu + \Delta_2 \, (p_2)_\mu (p_2)_\nu .
\end{align}
The form does not include graviton momenta $k^\mu$ for the sake of simplicity, as they can always be excluded by the conservation law $k + p_1 + p_2 =0$. This form includes all possible operators that can be constructed from the scalar momenta alone, so it is the most general expression. Functions $\alpha$, $\beta$, and $\Delta_i$ depend only on invariants constructed from all three momenta. Lastly, the operator can be put in the following more symmetric form:
\begin{align}
  \begin{split}
    i\,\Gamma_{\mu\nu} = ~& \alpha \,\eta_{\mu\nu} + \beta \left[ (p_1)_\mu (p_2)_\nu + (p_2)_\mu (p_1)_\nu \right]\\
    & + \cfrac{\Delta_1 + \Delta_2}{2}\,\left[ (p_1)_\mu (p_1)_\nu + (p_2)_\mu (p_2)_\nu \right] + \cfrac{\Delta_1 - \Delta_2}{2}\,\left[ (p_1)_\mu (p_1)_\nu - (p_2)_\mu (p_2)_\nu \right] .
  \end{split}
\end{align}
This form is explicitly symmetric with respect to index transposition $\mu\leftrightarrow\nu$ and scalar field momentum transmutation $p_1 \leftrightarrow p_2$.

We calculated the ultraviolet divergent part of the operator:
\begin{align}
  \begin{split}
    \alpha\Big|_\text{UV} & = i\,\pi^2\,\kappa^3\,\cfrac{1}{48}\,\cfrac{1}{\varepsilon_\text{UV}} \Bigg[ - 12\, m^4 + 2 m^2 \Big( 11\,k^2 + 3\, \left( p_1^2 + p_2^2 \right) \Big) - 2 \Big(k^2 + 2\,p_1^2 - p_2^2 \Big) \Big(k^2 - p_1^2 +2 \, p_2^2 \Big) \Bigg] ,\\
    \beta\Big|_\text{UV} &= i\,\pi^2\,\kappa^3\,\cfrac{1}{48}\,\cfrac{1}{\varepsilon_\text{UV}} \Bigg[ - 10\, m^2 + 8 \,k^2 + 7 \, \Big( p_1^2 + p_2^2 \Big) \Bigg] ,\\
    \Delta_1 \Big|_\text{UV} &= i\,\pi^2\,\kappa^3\,\cfrac{1}{48}\,\cfrac{1}{\varepsilon_\text{UV}} \Bigg[ - 16 \, m^2 + 12 \Big( k^2 - p_1^2 \Big) + 30 \, p_2^2 \Bigg] , \\
    \Delta_2 \Big|_\text{UV} &= i\,\pi^2\,\kappa^3\,\cfrac{1}{48}\,\cfrac{1}{\varepsilon_\text{UV}} \Bigg[ - 16 \, m^2 + 12 \Big( k^2 - p_2^2 \Big) + 30 \, p_1^2 \Bigg] .
  \end{split}
\end{align}
Here $\varepsilon_\text{UV}$ is the ultraviolet divergent factor. The functions present in the symmetric form of the operator read:
\begin{align}
  \begin{split}
    \cfrac{\Delta_1+\Delta_2}{2} &= i\,\pi^2\,\kappa^3\,\cfrac{1}{48}\,\cfrac{1}{\varepsilon_\text{UV}} \Bigg[ -16 \,m^2 + 12\, k^2 + 9 \big( p_1^2 +p_2^2 \big) \Bigg] \,, \\
    \cfrac{\Delta_1-\Delta_2}{2} &= i\,\pi^2\,\kappa^3\,\cfrac{1}{48}\,\cfrac{1}{\varepsilon_\text{UV}}\Bigg[- 21\, \big(p_1^2 - p_2^2 \big) \Bigg] \,.
  \end{split}
\end{align}

Lastly, the vertex operator can be expressed in terms of simpler operators. The following terms provide the minimal set of operators sufficient to describe the one-loop vertex operator:
\begin{align}
  \begin{split}
    \int d^4 x \sqrt{-g} \Bigg[ - \cfrac{m^2}{2} \,\phi^2 \Bigg] & \to - \cfrac{\kappa}{2} \, \eta_{\mu\nu} h^{\mu\nu}(k) \, \phi(p_1) \, \phi(p_2)\, ,\\
    \int d^4 x \sqrt{-g} \Bigg[ \cfrac12\, g^{\mu\nu} \,\nabla_\mu \phi \,\nabla_\nu\phi \Bigg] &\to - \cfrac{\kappa}{2} \, \Big[ \eta_{\mu\alpha} \eta_{\nu\beta} + \eta_{\mu\beta} \eta_{\nu\alpha} - \eta_{\mu\nu} \eta_{\alpha\beta} \Big] \, (p_1)^\alpha \, (p_2)^\beta \, h^{\mu\nu}(k)\,\phi(p_1)\, \phi(p_2) ,\\
    \int d^4 x \sqrt{-g} \Bigg[ R\, \phi^2 \Bigg] & \to -\kappa \, \Big[ k^2\, \eta_{\mu\nu} - k_\mu k_\nu \Big] \,h^{\mu\nu}(k) \,\phi(p_1) \,\phi(p_2), \\
    \int d^4 x \sqrt{-g} \Bigg[ R^{\mu\nu} \, \nabla_\mu \phi \, \nabla_\nu \phi \Bigg] &\to -\cfrac{\kappa}{4} \Bigg[ k_\mu k_\alpha \eta_{\nu\beta} + k_\mu k_\beta \eta_{\nu\alpha} + k_\nu k_\alpha \eta_{\mu\beta} + k_\nu k_\beta \eta_{\mu\alpha} \\
      &\hspace{60pt} - k^2 \left( \eta_{\mu\alpha}\eta_{\nu\beta} + \eta_{\mu\beta} \eta_{\nu\beta} \right) - 2 \,k_\mu k_\nu \,\eta_{\alpha\beta}\Bigg] h^{\mu\nu}(k) \,\phi(p_1) \,\phi(p_2) \,.
  \end{split}
\end{align}
The corresponding operators take the following form:
\begin{align}
  \begin{split}
    \mathcal{O}_1 &= -\cfrac12\,\kappa\,\eta_{\mu\nu} \,, \\
    \mathcal{O}_2 &= -\cfrac12\,\kappa\,\Big[ (p_1)_\mu (p_2)_\nu + (p_2)_\mu (p_1)_\nu - p_1\cdot p_2 \,\eta_{\mu\nu} \Big] \,, \\
    \mathcal{O}_3 &= \kappa\,\Big[ (p_1)_\mu (p_2)_\nu + (p_2)_\mu (p_1)_\nu + (p_1)_\mu (p_1)_\nu + (p_2)_\mu (p_2)_\nu  - (p_1 + p_2)^2 \,\eta_{\mu\nu} \Big] \,, \\
    \mathcal{O}_4 &= -\cfrac12\,\kappa\,\Big[ p_2^2\,(p_1)_\mu (p_1)_\nu + p_1^2 \, (p_2)_\mu (p_2)_\nu -\left\{ (p_1)_\mu (p_2)_\nu + (p_2)_\mu (p_1)_\nu\right\} \,p_1\cdot p_2 \Big]\,.
  \end{split}
\end{align}
The one-loop vertex operator is expressed as follows:
\begin{align}\label{Vertex_operator_UV_structure}
  \begin{split}
    i\,\Gamma_{\mu\nu} \Big|_\text{UV} \!= \cfrac{ i\,\pi^2 \kappa^2}{48}\,\cfrac{1}{\varepsilon_\text{UV}} \Bigg[ \Big\{ 24\,m^4 - 6\,m^2\! \left(3\,k^2 + p_1^2 + p_2^2  \right) + 47 \, k^2\! \left( p_1^2 + p_2^2\right) +  16\, p_1^2 \,p_2^2 - 10 \left( p_1^4 + p_2^4\right)  \Big\} \mathcal{O}_1\\
      - \Big\{ 12\,m^2 + 34\,k^2 - 4 \left( p_1^2 + p_2^2 \right) \Big\} \mathcal{O}_2 - \left\{ 16\,m^2 - 12 \left( k^2 - p_2^2 - p_2^2\right) \right\} \mathcal{O}_3 - 84\, \mathcal{O}_4 \Bigg] .
  \end{split}
\end{align}

\subsection{One-loop scattering amplitude}\label{Section_One-Loop_Scattering_Amplitude}

We use the following notations for the one-loop scalar four-point function:
\begin{align}
  \nonumber \\
  i\,\mathcal{M}(p_1,p_2,p_3,p_4,m_1,m_2) = \hspace{20pt}
  \begin{gathered}
    \begin{fmffile}{Diags/D}
      \begin{fmfgraph*}(50,50)
        \fmftop{T1,T2}
        \fmfbottom{B1,B2}
        \fmf{dashes}{B1,V1}
        \fmf{dashes}{B2,V2}
        \fmf{dashes}{T1,V1}
        \fmf{dashes}{T2,V2}
        \fmf{phantom,tension=2}{V1,V,V2}
        \fmfblob{30}{V}
        \fmfdot{V1,V2}
        \fmflabel{$p_1$}{B1}
        \fmflabel{$p_2$}{B2}
        \fmflabel{$p_3$}{T1}
        \fmflabel{$p_4$}{T2}
      \end{fmfgraph*}
    \end{fmffile}
  \end{gathered}
\end{align}
Unlike the previous case, it does not disappear when all momenta are on the shell. However, for the sake of generality, we will examine it with momenta beginning off-shell.

We have presented the complete amplitude and its components in the referenced repository \cite{latosh2023}. The file "\texttt{Scalar\_Four-Point.dat}" contains the entire amplitude in FeynCalc format. The remaining files contain the following contributions:
\begin{align}
  \begin{split}
    \\
    \begin{gathered}
      \begin{fmffile}{Diags/DD1}
        \begin{fmfgraph*}(30,30)
          \fmfbottom{B1,B2}
          \fmftop{T1,T2}
          \fmf{dashes}{B1,V1,T1}
          \fmf{dashes}{B2,V2,T2}
          \fmf{dbl_wiggly}{V1,V2}
          \fmfdot{V2}
          \fmfblob{10}{V1}
          \fmflabel{$p_1$}{B1}
          \fmflabel{$p_2$}{B2}
          \fmflabel{$p_3$}{T1}
          \fmflabel{$p_4$}{T2}
        \end{fmfgraph*}
      \end{fmffile}
    \end{gathered} \hspace{20pt} &\to \text{``\texttt{Scalar\_Four-Point\_Vertex\_Operator\_Contribution.dat}''},
    \\ \\
    \begin{gathered}
      \begin{fmffile}{Diags/DD2}
        \begin{fmfgraph*}(30,30)
          \fmfbottom{B1,B2}
          \fmftop{T1,T2}
          \fmf{phantom,tension=0.5}{L1,R1,R2,L2,L1}
          \fmf{dashes}{B1,L1}
          \fmf{dashes}{B2,R1}
          \fmf{dashes}{T1,L2}
          \fmf{dashes}{T2,R2}
          \fmfdot{L1,L2,R1,R2}
          \fmffreeze
          \fmf{dashes}{L1,L2}
          \fmf{dashes}{R1,R2}
          \fmf{dbl_wiggly}{L1,R1}
          \fmf{dbl_wiggly}{L2,R2}
          \fmflabel{$p_1$}{B1}
          \fmflabel{$p_2$}{B2}
          \fmflabel{$p_3$}{T1}
          \fmflabel{$p_4$}{T2}
        \end{fmfgraph*}
      \end{fmffile}
    \end{gathered} \hspace{20pt} &\to \text{``\texttt{Scalar\_Four-Point\_Box\_Operator.dat}''},
    \\ \\ 
    \begin{gathered}
      \begin{fmffile}{Diags/DD3}
        \begin{fmfgraph*}(30,30)
          \fmfbottom{B1,B2}
          \fmftop{T1,T2}
          \fmf{phantom,tension=.5}{V1,V0,V2}
          \fmf{phantom,tension=.2}{V1,V2}
          \fmf{dashes}{B2,V1}
          \fmf{dashes}{T2,V2}
          \fmf{dashes}{B1,V0,T1}
          \fmfdot{V0,V1,V2}
          \fmffreeze
          \fmf{dashes}{V1,V2}
          \fmf{dbl_wiggly}{V1,V0}
          \fmf{dbl_wiggly}{V2,V0}
          \fmflabel{$p_1$}{B1}
          \fmflabel{$p_2$}{B2}
          \fmflabel{$p_3$}{T1}
          \fmflabel{$p_4$}{T2}
        \end{fmfgraph*}
      \end{fmffile}
    \end{gathered} \hspace{20pt} &\to \text{``\texttt{Scalar\_Four-Point\_Wedge\_Operator.dat}''},
    \\ \\ 
    \begin{gathered}
      \begin{fmffile}{Diags/DD4}
        \begin{fmfgraph*}(30,30)
          \fmftop{T1,T2}
          \fmfbottom{B1,B2}
          \fmf{dashes}{B1,V1,T1}
          \fmf{dashes}{B2,V2,T2}
          \fmf{dbl_wiggly,right=1}{V1,V2,V1}
          \fmfdot{V1,V2}
          \fmflabel{$p_1$}{B1}
          \fmflabel{$p_2$}{B2}
          \fmflabel{$p_3$}{T1}
          \fmflabel{$p_4$}{T2}
        \end{fmfgraph*}
      \end{fmffile}
    \end{gathered} \hspace{20pt} &\to \text{``\texttt{Scalar\_Four-Point\_Bubble\_Operator.dat}''}. \\
  \end{split}
\end{align}
The files ending with ``\texttt{*\_Simple\_Form.dat}'' use the following generalisation of the Mandelstam variables. Lastly, files ending with ``\texttt{*\_UV.dat}'' contain only the ultraviolet divergent part of the corresponding operator.

The Mandelstam variables are usually defined for on-shell amplitudes, but extending them to off-shell amplitudes is straightforward. For an off-shell amplitude with four in-going momenta $p_i$ related by the conservation law $p_1 + p_2 + p_3 + p_4 =0$ we define the variables in the standard way:
\begin{align}
  s & \overset{\text{def}}{=} (p_1 + p_2)^2 \,, & t & \overset{\text{def}}{=} (p_1 + p_3)^2 \,, & u & \overset{\text{def}}{=} (p_1 + p_4)^4 \,.
\end{align}
Because the momenta are begin off-shell, the new variables enjoy the following relation:
\begin{align}
  s + t + u = p_1^2 + p_2^2 + p_3^2 + p_4^2 \,.
\end{align}
These variables provide a way to express the scattering amplitude through the following scalar quantities alone, which improves calculation efficiency:
\begin{align}
  s\,, t\,, u\,, p_1^2\,, p_2^2\,, p_3^2\,, p_4^2\,, m_1\,, m_2.
\end{align}

The amplitude contains the following Passarino-Veltman integrals:
\begin{align}
  \begin{split}\
    & A_0 (m_1^2) \,, A_0 (m_2^2) \,, B_0 (t,0,0) \,, B_0 (t,m_1^2,m_1^2) \,, B_0 (t,m_2^2,m_2^2) \,, B_0 (s,m_1^2,m_2^2) \,, B_0 (u,m_1^2,m_2^2) \,,\\
    & B_0 (p_1^2,0,m_1^2) \,, B_0 (p_3^2,0,m_1^2) \,, B_0 (p_2^2,0,m_2^2) \,, B_0 (p_4^2,0,m_2^2) \,, C_0 (p_1^2,p_2^2,s,m_1^2,0,m_2^2) \,,\\
    & C_0 (s,p_3^2,p_4^2,m_2^2,m_1^2,0)\,, C_0(p_1^2,p_4^2,u,m_1^2,0,m_2^2) \,, C_0 (u,p_3^2,p_2^2,m_2^2,m_1^2,0) \,, \\
    & C_0 (p_1^2,p_3^2,t,0,m_1^2,0) \,, C_0 (p_1^2,p_3^2,t,m_1^2,0,m_1^2) \,, C_0 (p_1^2,t,p_3^2,m_1^2,0,0) \,, \\
    & C_0 (p_2^2,p_4^2,t,0,m_2^2,0) \,, C_0 (p_2^2,p_4^2,t,m_2^2,0,m_2^2) \,, C_0 (p_2^2,t,p_4^2,m_2^2,0,0) \,, \\
    & D_0 (p_1^2,p_2^2,p_4^2,p_3^2,s,t,m_1^2,0,m_2^2,0) \,, D_0 (p_1^2,p_4^2,p_2^2,p_3^2,u,t,m_1^2,0,m_2^2,0) \,.
  \end{split}
\end{align}
If momenta are off-shell, then the amplitude has no infrared divergences. In the most interesting physical case, the amplitude receives the standard infrared divergences when all the momenta are on-shell. The corresponding part of the operator reads:
\begin{align}
  \begin{split}
    i\, \mathcal{M} \Big|_\text{IR} = &\cfrac{i\,\pi^2\,\kappa^4}{16\,\varepsilon_\text{IR}} \,\cfrac{1}{t}\,\Bigg[ \cfrac{ 2\,m_2^4 - 4\, m_2^2 \, t + t^2 }{\left[ t\, (t - 4\, m_1^2)\right]^{3/2}} \,\left[ m_1^4 - m_1^2 ( t + 2\, u) + ( m_2^2 - u ) (m_2^2 - t - u )  \right] \\
      & \hspace{45pt} \times \ln \frac{\sqrt{t ( t - 4\, m_2^2) } + 2 \,m_2^2 - t}{2\, m_2^2} + \left[m_1 \leftrightarrow m_2 \right] \Bigg]\\
    & + \cfrac{i\,\pi^2\,\kappa^4}{8\,\varepsilon_\text{IR}} \,\cfrac{1}{t} \Bigg[ \cfrac{\left[ m_1^4 - 2 \,m_2^2 \, s + ( m_2^2 - s )^2 \right] \left[ m_1^4 - m_1^2 ( t + 2\, u) + ( m_2^2 - u ) ( m_2^2 - t - u ) \right]}{ \sqrt{( t + u -s - 4\, m_1 \,m_2 ) (t + u -s + 4\, m_1 \,m_2)} } \\
      & \hspace{55pt} \times\ln \frac{m_1^2 + m_2^2 - s +\sqrt{ m_1^4 - 2\, m_1^2 ( m_2^2 + s ) + ( m_2^2 - s )^2} }{2\,m_1\,m_2} + \left[ s \leftrightarrow u \right] \Bigg].
  \end{split}
\end{align}

The amplitude also has ultraviolet divergences. The corresponding operator reads:
\begin{align}
  \begin{split}
    i\,\mathcal{M} \Big|_\text{UV} = & \cfrac{i\,\pi^2\,\kappa^4}{192\,\varepsilon_\text{UV} }\,\cfrac{1}{t} \Bigg[ 96\, m_1^4 m_2^2 - 24\, m_1^4 p_2^2 - 24\, m_1^4 p_4^2 + 96 \, m_1^2 m_2^4 - 36\, m_1^2 m_2^2 s - 36\, m_1^2 m_2^2 u - 16 \, m_1^2 p_2^4 \\
      & \hspace{45pt} + 12\, m_1^2 p_2^2 p_3^2 - 4\, m_1^2 p_2^2 p_4^2 + 6\, m_1^2 p_2^2 s + 6\, m_1^2 p_2^2 u + 12 \, m_1^2 p_3^4 + 12\, m_1^2 p_3^2 p_4^2 - 12\, m_1^2 p_3^2 s \\
      & \hspace{45pt} - 12\, m_1^2 p_3^2 u - 16\, m_1^2 p_4^4 + 6\, m_1^2 p_4^2 s + 6\, m_1^2 p_4^2 u + 12\, m_1^2 s u - 24\, m_2^4 p_1^2 - 24\, m_2^4 p_3^2 \\
      & \hspace{45pt} + 24\, m_2^4 t - 16\, m_2^2 p_2^4 - 28 \, m_2^2 p_2^2 p_3^2 - 44 \, m_2^2 p_2^2 p_4^2 + 26\, m_2^2 p_2^2 s + 26\, m_2^2 p_2^2 u- 28\, m_2^2 p_3^4 \\
      & \hspace{45pt} - 28 \, m_2^2 p_3^2 p_4^2 + 28\, m_2^2 p_3^2 s + 28 \, m_2^2 p_3^2 u - 16\, m_2^2 p_4^4 + 26 \, m_2^2 p_4^2 s + 26 \, m_2^2 p_4^2 u - 10\, m_2^2 s^2\\
      & \hspace{45pt} - 8\, m_2^2 s u - 10 \, m_2^2 u^2 + 72\, p_2^4 p_3^2 + 72\, p_2^4 p_4^2 - 61\, p_2^4 s + 23 \, p_2^4 u + 72 \,p_2^2 p_3^4 + 144 \, p_2^2 p_3^2 p_4^2 \\
      & \hspace{45pt} - 160 \, p_2^2 p_3^2 s + 8\, p_2^2 p_3^2 u + 72 \, p_2^2 p_4^4 - 106 \, p_2^2 p_4^2 s - 106\, p_2^2 p_4^2 u + 61 \, p_2^2 s^2 + 38 \, p_2^2 s u - 23\, p_2^2 u^2\\
      & \hspace{45pt} + 72 \, p_3^4 p_4^2 - 4\, p_3^4 s - 4\, p_3^4 u + 72\, p_3^2 p_4^4 + 8\, p_3^2 p_4^2 s - 160\, p_3^2 p_4^2 u + 4\, p_3^2 s^2 + 8\, p_3^2 s u + 4\, p_3^2 u^2 \\
      & \hspace{45pt} + 23 \, p_4^4 s - 61 \, p_4^4 u - 23\, p_4^2 s^2 + 38 \, p_4^2 s u + 61\, p_4^2 u^2 - 4 \, s^2 u - 4\, s u^2 \Bigg]\\
    & +\cfrac{i\,\pi^2\,\kappa^4}{192\,\varepsilon_\text{UV} } \Bigg[ 24\, m_1^4 - 228 \, m_1^2 m_2^2 - 6\, m_1^2 p_1^2 + 38\, m_1^2 p_2^2 - 18 \, m_1^2 p_3^2 + 38 \, m_1^2 p_4^2 - 22 \, m_1^2 t + 12 \, m_2^2 p_1^2\\
      & \hspace{45pt} + 40\, m_2^2 p_3^2 - 12\, m_2^2 t - 13\, p_1^2 p_2^2 - 28 \, p_1^2 p_3^2 - 13 \, p_1^2 p_4^2 + 13\, p_1^2 s + 13 \, p_1^2 u - 85 \, p_2^2 p_3^2 \\
      & \hspace{45pt} - 100 \, p_2^2 p_4^2 + 74 \, p_2^2 s - 10 \, p_2^2 u - 85 \, p_3^2 p_4^2 + 17 \, p_3^2 s + 17 \, p_3^2 u - 10 \, p_4^2 s + 74 \, p_4^2 u \\
      & \hspace{45pt} + 10\, s^2 - 48\, s u + 10\, u^2 \Bigg].
  \end{split}
\end{align}
That amplitude spawns the following set of operators ($i=1,2$):
\begin{align}
  \begin{split}
    m_i^4\,\phi_1^2\,\phi_2^2 \,, ~ m_i^2 \square \phi_1^2 \phi_2^2 \,, ~ m_i^2 \phi_1^2 \,\square \phi_2^2 \,,~ m_i^2 \pd \phi_1^2 \pd \phi_2^2 \,, ~\square \phi_1^2 \, \square \phi_2^2\, , ~ \pd \phi_1^2 \, \square \pd \phi_2^2 \, , ~ \square \pd \phi_1^2 \, \pd \phi_2^2 \, , ~ \pd_\mu \pd_\nu \phi_1^2 \, \pd^\mu \pd^\nu \phi_2^2 \, .
  \end{split}
\end{align}

Once again, let us highlight that the complete amplitude is published separately \cite{latosh2023} in open access. In the FeynCalc notation, it takes more than 600 MB. The simplified form given in terms of the generalised Mandelstam variables takes about 64 MB. In both cases, it is impossible to present it in print. 

\subsection{Low-energy limit}

Let us proceed with the low-energy limit of the theory. To our knowledge, the limit was first studied in \cite{Donoghue:1994dn,Burgess:2003jk}. These and many other results help develop a method of classical potentials computation in compact binaries \cite{Bern:2019crd,Bern:2020buy,Chung:2018kqs}. These methods are based on modern tools for amplitude calculations and rely heavily on on-shall scattering amplitudes. As defined above, the low-energy limit is when all external momenta approach zero. Let us consider amplitudes \eqref{Amplitude_one-loop} as a function of external momenta. In full accordance with the BPHZ theorem \cite{Bogoliubov:1957gp,Hepp:1966eg,Zimmermann:1969jj}, all ultraviolet divergent terms are analytic functions of external momenta. Therefore, in the low-energy limit, all divergences are subdominant. On the contrary, ultraviolet finite terms of \eqref{Amplitude_one-loop} contain non-analytic functions of external momenta that become dominant in the low-energy limit and provide the leading order corrections to the scattering cross-section.

All calculations were performed on the mass shell with the standard Mandelstam variables to extract the low-energy limit. Secondly, all loop integrals that did not contain the Mandelstam variables were omitted as they do not contain non-analytic functions of external momenta. Other loop integrals were calculated with the packages mentioned above \cite{Patel:2015tea,Patel:2016fam,Shtabovenko:2016whf} and were replaced by the corresponding asymptotic expressions in $p_\text{cm} \to 0$ limit. We summarise the corresponding asymptotic expressions in Appendix \ref{appendix_asymptotic}. Lastly, the overall expression was evaluated with the standard FeynCalc tools in $d=4$, and the corresponding asymptotically expression in the same limit was obtained.

The leading order expression for the amplitudes \eqref{Amplitude_one-loop} in the IR limit reads:
\begin{align}
  \mathcal{M}_\text{IR} = i\, G\, 4\,\pi\, \cfrac{m_1^2\,m_2^2}{p_\text{cm}^2\, \sin^2\cfrac{\theta}{2}} + i\, G^2\, 64\,\pi^2 \, \cfrac{m_1^2\,m_2^2\,(m_1+m_2)}{p_\text{cm}\,\sin\cfrac{\theta}{2}} - G^2\, 128\,\pi^5 \, \cfrac{ m_1^4\, m_2^4\, \ln p_\text{cm}}{(m_1+m_2) \, p_\text{cm}^3\,\sin^2\cfrac{\theta}{2}} + \mathcal{O}\left(G^3\right).
\end{align}
It results in the following expression of the differential cross-section
\begin{align}\label{the_main_result}
  \left. \cfrac{d\sigma}{d\Omega} \right|_\text{IR limit}  = \cfrac{1}{4} \, \left(G\, \mu \, m_1 \, m_2\right)^2 \cfrac{1}{p_\text{cm}^4 \, \sin^4\cfrac{\theta}{2}} \left[ 1 + 32\,\pi^5\,G\, p_\text{cm}\,(m_1+m_2) \, \sin\cfrac{\theta}{2} + \mathcal{O}\left( G^2\right) \right].
\end{align}
Firstly, this result is in agreement with previous non-relativistic and relativistic calculations. Secondly, it provides explicit corrections generated at the one-loop level. Lastly, the expression is given in terms of variables calculated in the centre of the mass frame while all particles are fixed on the mass shell. Because of this, the expression is gauge invariant and can be transferred to any other given frame. This makes it possible to subject it to direct empirical verification. We discuss the physical implications of this result in the next section.

\section{Discussion and conclusions}\label{Section_conclusion}

This paper studies the gravitational scattering of two scalar particles of different masses in perturbative quantum gravity. The perturbative approach to quantum gravity produces a theory that is non-renormalisable in the standard way but helps to extract meaningful information from scattering amplitudes. In turn, the scattering of scalar particles is the simplest example of a process that can be studied explicitly. The corresponding amplitude contains the least possible number of terms, as it does not involve Lorentz and spinor indices. The recently developed package FeynGrav provides a tool to explicitly calculate the amplitude and all the involved operators.

We proved that at the tree level, the perturbative quantum gravity completely agrees with the classical results. First and foremost, we investigated the consistency with the classical results. In Section \ref{Section_non-relativistic_scattering}, we generalised the Rutherford formula for gravitational scattering cross section \eqref{differential_cross_section_NR}. In Section \ref{Section_relativistic_scattering}, we proposed a way to obtain a suitable generalisation of the scattering for the relativistic case. The most general case of classical relativistic gravitational scattering is given by the Einstein-Infeld-Hoffman equations, but it involves non-potential terms. We studied the case of scattering of a probe particle on the Schwarzschild black hole. The case is similar to scattering on a rigidly fixed body, and all non-potential interactions are suppressed by the probe particle/black hole mass hierarchy. Equation \eqref{scattering_angle_relativistic_case} expresses the scattering angle as a function of the impact parameter, but it is impossible to solve it for the impact parameter analytically. With the help of numerical calculations, we established an analytic approximation for the scattering cross section \eqref{BH_cross_section_small_angle}, which agreed with the Rutherford formula. Lastly, we calculated the gravitational scattering cross section of two scalars of different mass at the tree level in Section \ref{Section_perturbative_quantum_gravity}. We obtained an explicit analytic expression \eqref{cross_section_with_factors}, which agrees with relativistic and non-relativistic classical results.

We extended the investigation beyond the tree level in Section \ref{Secition_one-loop}. The explicit expression for one-loop operators contains multiple terms, so presenting them in print is impossible and impractical. We published them in open access separately \cite{latosh2023}. In Section \ref{Section_one-loop_vertex_operator}, we studied the one-loop operator describing the interaction between a scalar field and gravity. The operator contains the standard infrared divergences in massive \eqref{Vertex_operator_IR_1} and massless \eqref{Vertex_operator_IR_2} cases. The structure of its ultraviolet divergences is given explicitly by \eqref{Vertex_operator_UV_structure}. In Section \ref{Section_One-Loop_Scattering_Amplitude}, we discuss the corresponding scattering amplitude. We explicitly present its infrared and ultraviolet divergences. Lastly, we study the low-energy limit of the process. This approach was widely used to reconstruct a non-relativistic gravitational potential \cite{Vanhove:2021zel,Donoghue:1994dn,Donoghue:2001qc,BjerrumBohr:2002kt,Bjerrum-Bohr:2002fji,Bjerrum-Bohr:2013bxa,Bjerrum-Bohr:2014zsa,Bjerrum-Bohr:2016hpa,Bjerrum-Bohr:2018xdl,Bjerrum-Bohr:2021vuf}. However, this method has a significant disadvantage, as the potential is gauge dependent, so it is not defined uniquely, and cannot be directly probed in an experiment. In this paper, we provided an explicit example of quantum calculations that provide the leading contribution to a scattering cross-section in the low-energy limit \eqref{the_main_result}. In contrast to the above results, this scattering cross-section is gauge-independent. Moreover, it is given in terms of the centre of mass variables so that it can be transferred to any frame. Consequently, it is possible to subject it to direct empirical verification.

It must be noted that additional powers of the Newton constant suppress the discussed loop corrections. This behaviour is typical for any perturbative calculations within quantum field theory, as higher powers of couplings suppress higher-order perturbative corrections. Approaches that may reduce such a suppression lie far beyond the standard perturbation theory. Nonetheless, as the earlier papers show, scattering amplitudes account for both quantum and classical effects (see for the most detailed discussion \cite{Bjerrum-Bohr:2021vuf}). Consequently, it is possible to compare scattering cross-sections with empirical data directly.

Results obtained in the present paper point towards a few promising ways to extend the study. First and foremost, this paper provides an explicit example of FeynGrav usage for calculations within the perturbative quantum gravity. In the latest version, FeynGrav contains Feynman rules sufficient to describe quantum gravity within the standard model \cite{Latosh:2022ydd,Latosh:2023zsi}. 

Secondly, the gravitational light scattering on a massive object is the most prospective object of further study. The Event Horizon Telescope can directly probe contemporary gravitational light scattering on a black hole \cite{EventHorizonTelescope:2019dse}. It is challenging to comprehensively describe a black hole and its gravitational interaction with an external photon within the perturbative quantum gravity. The scattering of light on a massive object captures the leading contribution from gravitational interaction, regardless of whether it is scalar, vector, or fermion. Scattering amplitudes have already undergone similar calculations, resulting in predictions for deflection angles. \cite{Bjerrum-Bohr:2014zsa,Bjerrum-Bohr:2016hpa,Bai:2016ivl,Chi:2019owc}. FeynGrav provides a new way to extend calculations beyond on-shell amplitudes and massless articles.

\section*{Acknowledgement}
The work (BL) was supported by the Foundation for the Advancement of Theoretical Physics and Mathematics "BASIS".

\appendix

\section{Asymptotic expressions for loop integrals}\label{appendix_asymptotic}

In the paper, we use asymptotic values of loop integral in the IR limit of the theory. Here, we present a summary of their asymptotic behaviour. 

Firstly, integrals that do not contain Mandelstam variables are irrelevant to the IR limit because they cannot contain any non-analytic functions of external moments. Within the studied one-loop amplitudes, this applies only to the following integrals:
\begin{align}
  \begin{split}
    & A_0 \left( m_1^2 \right),  A_0 \left( m_1^2 \right)  \to 0\, , \\
    & B_0 \left( m_1^2, 0, m_1^2 \right), B_0 \left( m_2^2, 0, m_2^2 \right) \to 0 .
  \end{split}
\end{align}
Secondly, some loop integrals that do contain Mandelstam variables are irrelevant for the IR limit because they admit a smooth finite limit. The following integrals have this feature and appear in the discussed amplitudes:
\begin{align}
  \begin{split}
    & B_0 \left( s, m_1^2, m_2^2 \right), B_0 \left( t, m_1^2, m_1^2 \right), B_0 \left( t, m_2^2, m_2^2 \right) \to 0 \,,\\
    & C_0 \left( m_1^2, m_1^2, t, m_1^2, 0, m_1^2 \right), C_0 \left( m_2^2, m_2^2, t, m_2^2, 0, m_2^2 \right) \to 0\,.
  \end{split}
\end{align}
For illustration, in detail, let us consider $B_0 (t, m_1^2, m_1^2)$. For the given kinematics, we can express $t$ in terms of the centre of mass momentum $p_\text{cm}$ and the centre of mass scattering angle $\theta$. This provides us with the following expression:
\begin{align}
  \begin{split}
    B_0 \left( t, m_1^2, m_1^2 \right) =& \cfrac{1}{d-4} - \gamma + 2 + \ln \cfrac{\mu^2}{\pi\,m_1^2} \\
    &- \sqrt{1+ \cfrac{m_1^2}{p_\text{cm}^2\,\sin^2\frac{\theta}{2}}}\, \ln \left[ 1 + 2\, \cfrac{p_\text{cm}^2}{m_1^2} \,\sin^2\frac{\theta}{2} + 2 \, \cfrac{p_\text{cm}}{m} \,\sin\frac{\theta}{2} \sqrt{ 1 + \cfrac{p_\text{cm}^2}{m_1^2} \, \sin^2\frac{\theta}{2} }  \right].
  \end{split}
\end{align}
In the IR limit, we shall take $p_\text{cm} \to 0$, but this expression remains finite. Therefore, it will not provide the dominant contribution to any matrix element.

Amplitudes studied in this paper admit only seven loop integrals, which will provide a dominant contribution within the IR limit and that define the low energy structure of the theory:
\begin{align}
  \begin{split}
    & B_0 \left( t, 0, 0 \right) \to -\ln p^2 \, ,\\
    & C_0 \left( m_1^2, m_1^2, t, 0, m_1^2, 0 \right) \to - \cfrac{\pi^2}{4\,m_1^2\,p\, \sin \frac{\theta}{2}} \,, \\
    & C_0 \left( m_2^2, t, m_2^2, m_2^2, 0, 0 \right), C_0 \left( m_2^2, m_2^2, t, 0, m_2^2, 0 \right) \to - \cfrac{\pi^2}{4\,m_2^2\,p\, \sin \frac{\theta}{2}} \,, \\
    & C_0 \left( m_1^2, m_2^2, s, m_1^2, 0, m_2^2 \right), C_0 \left( m_1^2, s, m_2^2, 0, m_1^2,  m_2^2 \right) \to -\cfrac{i\,\pi\,\ln p}{p\,(m_1+m_2)} \,,\\
    & D_0 \left( m_1^2, m_1^2, m_2^2, m_2^2, t, s, 0, m_1^2, 0, m_2^2 \right) \to \cfrac{i\,\pi\,\ln p}{2\,p^3\,(m_1+m_2)\,\sin^2\frac{\theta}{2}}\,.
  \end{split}
\end{align}

\bibliographystyle{unsrt}
\bibliography{oTBGSwPG.bib}

\begin{thebibliography}{10}

\bibitem{tHooft:1974toh}
Gerard 't~Hooft and M.~J.~G. Veltman.
\newblock {One loop divergencies in the theory of gravitation}.
\newblock {\em Ann. Inst. H. Poincare Phys. Theor.}, A20:69--94, 1974.

\bibitem{Deser:1974cz}
Stanley Deser and P.~van Nieuwenhuizen.
\newblock {One Loop Divergences of Quantized Einstein-Maxwell Fields}.
\newblock {\em Phys. Rev. D}, 10:401, 1974.

\bibitem{Deser:1974cy}
Stanley Deser and P.~van Nieuwenhuizen.
\newblock {Nonrenormalizability of the Quantized Dirac-Einstein System}.
\newblock {\em Phys. Rev. D}, 10:411, 1974.

\bibitem{Goroff:1985sz}
Marc~H. Goroff and Augusto Sagnotti.
\newblock {Quantum gravity at two loops}.
\newblock {\em Phys. Lett. B}, 160:81--86, 1985.

\bibitem{Latosh:2020jyq}
Boris Latosh.
\newblock {One-loop effective scalar-tensor gravity}.
\newblock {\em Eur. Phys. J. C}, 80(9):845, 2020.

\bibitem{Horndeski:1974wa}
Gregory~Walter Horndeski.
\newblock {Second-order scalar-tensor field equations in a four-dimensional
  space}.
\newblock {\em Int. J. Theor. Phys.}, 10:363--384, 1974.

\bibitem{Kobayashi:2011nu}
Tsutomu Kobayashi, Masahide Yamaguchi, and Jun'ichi Yokoyama.
\newblock {Generalized G-inflation: Inflation with the most general
  second-order field equations}.
\newblock {\em Prog. Theor. Phys.}, 126:511--529, 2011.

\bibitem{Kase:2018aps}
Ryotaro Kase and Shinji Tsujikawa.
\newblock {Dark energy in Horndeski theories after GW170817: A review}.
\newblock {\em Int. J. Mod. Phys. D}, 28(05):1942005, 2019.

\bibitem{LIGOScientific:2017zic}
B.~P. Abbott et~al.
\newblock {Gravitational Waves and Gamma-rays from a Binary Neutron Star
  Merger: GW170817 and GRB 170817A}.
\newblock {\em Astrophys. J. Lett.}, 848(2):L13, 2017.

\bibitem{Ezquiaga:2017ekz}
Jose~Mar\'\i{}a Ezquiaga and Miguel Zumalac\'arregui.
\newblock {Dark Energy After GW170817: Dead Ends and the Road Ahead}.
\newblock {\em Phys. Rev. Lett.}, 119(25):251304, 2017.

\bibitem{Dixon:1996wi}
Lance~J. Dixon.
\newblock {Calculating scattering amplitudes efficiently}.
\newblock In {\em {Theoretical Advanced Study Institute in Elementary Particle
  Physics (TASI 95): QCD and Beyond}}, pages 539--584, 1 1996.

\bibitem{Elvang:2013cua}
Henriette Elvang and Yu-tin Huang.
\newblock {Scattering Amplitudes}.
\newblock 8 2013.

\bibitem{Vanhove:2021zel}
Pierre Vanhove.
\newblock {$S$-matrix approach to general gravity and beyond}.
\newblock In {\em {55th Rencontres de Moriond on QCD and High Energy
  Interactions}}, 4 2021.

\bibitem{Travaglini:2022uwo}
Gabriele Travaglini et~al.
\newblock {The SAGEX Review on Scattering Amplitudes}.
\newblock 3 2022.

\bibitem{Arkani-Hamed:2017jhn}
Nima Arkani-Hamed, Tzu-Chen Huang, and Yu-tin Huang.
\newblock {Scattering amplitudes for all masses and spins}.
\newblock {\em JHEP}, 11:070, 2021.

\bibitem{Latosh:2022ydd}
Boris Latosh.
\newblock {FeynGrav: FeynCalc extension for gravity amplitudes}.
\newblock {\em Class. Quant. Grav.}, 39(16):165006, 2022.

\bibitem{Latosh:2023zsi}
Boris Latosh.
\newblock {FeynGrav 2.0}.
\newblock {\em Comput. Phys. Commun.}, 292:108871, 2023.

\bibitem{Shtabovenko:2016sxi}
Vladyslav Shtabovenko, Rolf Mertig, and Frederik Orellana.
\newblock {New Developments in FeynCalc 9.0}.
\newblock {\em Comput. Phys. Commun.}, 207:432--444, 2016.

\bibitem{Shtabovenko:2020gxv}
Vladyslav Shtabovenko, Rolf Mertig, and Frederik Orellana.
\newblock {FeynCalc 9.3: New features and improvements}.
\newblock {\em Comput. Phys. Commun.}, 256:107478, 2020.

\bibitem{Shtabovenko:2016whf}
Vladyslav Shtabovenko.
\newblock {FeynHelpers: Connecting FeynCalc to FIRE and Package-X}.
\newblock {\em Comput. Phys. Commun.}, 218:48--65, 2017.

\bibitem{Patel:2015tea}
Hiren~H. Patel.
\newblock {Package-X: A Mathematica package for the analytic calculation of
  one-loop integrals}.
\newblock {\em Comput. Phys. Commun.}, 197:276--290, 2015.

\bibitem{Einstein:1938yz}
Albert Einstein, L.~Infeld, and B.~Hoffmann.
\newblock {The Gravitational equations and the problem of motion}.
\newblock {\em Annals Math.}, 39:65--100, 1938.

\bibitem{Einstein:1940mt}
Albert Einstein and L.~Infeld.
\newblock {The Gravitational equations and the problem of motion. 2.}
\newblock {\em Annals Math.}, 41:455--464, 1940.

\bibitem{Straumann:2013spu}
Norbert Straumann.
\newblock {\em {General Relativity}}.
\newblock Graduate Texts in Physics. Springer, Dordrecht, 2013.

\bibitem{Patel:2016fam}
Hiren~H. Patel.
\newblock {Package-X 2.0: A Mathematica package for the analytic calculation of
  one-loop integrals}.
\newblock {\em Comput. Phys. Commun.}, 218:66--70, 2017.

\bibitem{Donoghue:1994dn}
John~F. Donoghue.
\newblock {General relativity as an effective field theory: The leading quantum
  corrections}.
\newblock {\em Phys. Rev.}, D50:3874--3888, 1994.

\bibitem{Donoghue:2001qc}
John~F. Donoghue, Barry~R. Holstein, Bj\"orn Garbrecht, and Thomas Konstandin.
\newblock {Quantum corrections to the Reissner-Nordstr\"om and Kerr-Newman
  metrics}.
\newblock {\em Phys. Lett. B}, 529:132--142, 2002.
\newblock [Erratum: Phys.Lett.B 612, 311--312 (2005)].

\bibitem{BjerrumBohr:2002kt}
N.~E.~J Bjerrum-Bohr, John~F. Donoghue, and Barry~R. Holstein.
\newblock {Quantum gravitational corrections to the nonrelativistic scattering
  potential of two masses}.
\newblock {\em Phys. Rev.}, D67:084033, 2003.
\newblock [Erratum: Phys. Rev.D71,069903(2005)].

\bibitem{Bjerrum-Bohr:2002fji}
Niels Emil~Jannik Bjerrum-Bohr, John~F. Donoghue, and Barry~R. Holstein.
\newblock {Quantum corrections to the Schwarzschild and Kerr metrics}.
\newblock {\em Phys. Rev. D}, 68:084005, 2003.
\newblock [Erratum: Phys.Rev.D 71, 069904 (2005)].

\bibitem{Bjerrum-Bohr:2013bxa}
N.~E.~J. Bjerrum-Bohr, John~F. Donoghue, and Pierre Vanhove.
\newblock {On-shell Techniques and Universal Results in Quantum Gravity}.
\newblock {\em JHEP}, 02:111, 2014.

\bibitem{Bjerrum-Bohr:2014zsa}
N.~E.~J. Bjerrum-Bohr, John~F. Donoghue, Barry~R. Holstein, Ludovic Planté,
  and Pierre Vanhove.
\newblock {Bending of Light in Quantum Gravity}.
\newblock {\em Phys. Rev. Lett.}, 114(6):061301, 2015.

\bibitem{Bjerrum-Bohr:2016hpa}
N.~E.~J. Bjerrum-Bohr, John~F. Donoghue, Barry~R. Holstein, Ludovic Plante, and
  Pierre Vanhove.
\newblock {Light-like Scattering in Quantum Gravity}.
\newblock {\em JHEP}, 11:117, 2016.

\bibitem{Bjerrum-Bohr:2018xdl}
N.~E.~J. Bjerrum-Bohr, Poul~H. Damgaard, Guido Festuccia, Ludovic Plant\'e, and
  Pierre Vanhove.
\newblock {General Relativity from Scattering Amplitudes}.
\newblock {\em Phys. Rev. Lett.}, 121(17):171601, 2018.

\bibitem{Bjerrum-Bohr:2021vuf}
N.~Emil~J. Bjerrum-Bohr, Poul~H. Damgaard, Ludovic Plant\'e, and Pierre
  Vanhove.
\newblock {Classical gravity from loop amplitudes}.
\newblock {\em Phys. Rev. D}, 104(2):026009, 2021.

\bibitem{Bogoliubov:1957gp}
N.~N. Bogoliubov and O.~S. Parasiuk.
\newblock {On the Multiplication of the causal function in the quantum theory
  of fields}.
\newblock {\em Acta Math.}, 97:227--266, 1957.

\bibitem{Hepp:1966eg}
Klaus Hepp.
\newblock {Proof of the Bogolyubov-Parasiuk theorem on renormalization}.
\newblock {\em Commun. Math. Phys.}, 2:301--326, 1966.

\bibitem{Zimmermann:1969jj}
W.~Zimmermann.
\newblock {Convergence of Bogolyubov's method of renormalization in momentum
  space}.
\newblock {\em Commun. Math. Phys.}, 15:208--234, 1969.

\bibitem{Battista:2014oba}
Emmanuele Battista and Giampiero Esposito.
\newblock {Restricted three-body problem in effective-field-theory models of
  gravity}.
\newblock {\em Phys. Rev. D}, 89(8):084030, 2014.

\bibitem{Battista:2014ija}
Emmanuele Battista and Giampiero Esposito.
\newblock {Full three-body problem in effective-field-theory models of
  gravity}.
\newblock {\em Phys. Rev. D}, 90(8):084010, 2014.
\newblock [Erratum: Phys.Rev.D 93, 049901 (2016)].

\bibitem{Battista:2015zta}
Emmanuele Battista, Simone Dell'Agnello, Giampiero Esposito, and Jules Simo.
\newblock {Quantum effects on Lagrangian points and displaced periodic orbits
  in the Earth-Moon system}.
\newblock {\em Phys. Rev. D}, 91:084041, 2015.
\newblock [Erratum: Phys.Rev.D 93, 049902 (2016)].

\bibitem{Battista:2015wwa}
Emmanuele Battista, Simone Dell'Agnello, Giampiero Esposito, Luciano Di~Fiore,
  Jules Simo, and Aniello Grado.
\newblock {Earth-moon Lagrangian points as a test bed for general relativity
  and effective field theories of gravity}.
\newblock {\em Phys. Rev. D}, 92:064045, 2015.
\newblock [Erratum: Phys.Rev.D 93, 109904 (2016)].

\bibitem{Battista:2017xlm}
Emmanuele Battista, Angelo Tartaglia, Giampiero Esposito, David Lucchesi,
  Matteo~Luca Ruggiero, Pavol Valko, Simone Dell'Agnello, Luciano Di~Fiore,
  Jules Simo, and Aniello Grado.
\newblock {Quantum time delay in the gravitational field of a rotating mass}.
\newblock {\em Class. Quant. Grav.}, 34(16):165008, 2017.

\bibitem{Tartaglia:2018bjc}
Angelo Tartaglia, Giampiero Esposito, Emmanuele Battista, Simone Dell'Agnello,
  and Bin Wang.
\newblock {Looking for a new test of general relativity in the solar system}.
\newblock {\em Mod. Phys. Lett. A}, 33(24):1850136, 2018.

\bibitem{Battista:2020qqp}
Emmanuele Battista, Giampiero Esposito, and Angelo Tartaglia.
\newblock {An effective-gravity perspective on the
  Sun\textendash{}Jupiter\textendash{}comet three-body system}.
\newblock {\em Int. J. Geom. Meth. Mod. Phys.}, 17(11):2050168, 2020.

\bibitem{Landau1976Mechanics}
L.~D. Landau and E.~M. Lifshitz.
\newblock {\em Mechanics, Third Edition: Volume 1 (Course of Theoretical
  Physics)}.
\newblock Butterworth-Heinemann, 3 edition, 1976.

\bibitem{Arnold1997mathematical}
V.I. Arnold.
\newblock {\em Mathematical Methods of Classical Mechanics}.
\newblock Graduate Texts in Mathematics. Springer New York, 1997.

\bibitem{goldstein2002classical}
H.~Goldstein, C.P. Poole, and J.L. Safko.
\newblock {\em Classical Mechanics}.
\newblock Addison Wesley, 2002.

\bibitem{Chandrasekhar:1985kt}
Subrahmanyan Chandrasekhar.
\newblock {\em {The mathematical theory of black holes}}.
\newblock 1985.

\bibitem{Perlick:2021aok}
Volker Perlick and Oleg~Yu. Tsupko.
\newblock {Calculating black hole shadows: Review of analytical studies}.
\newblock {\em Phys. Rept.}, 947:2190, 2022.

\bibitem{Collins:1973xf}
P.~A. Collins, Robert Delbourgo, and R.~M. Williams.
\newblock {On the elastic Schwarzschild scattering cross-section}.
\newblock {\em J. Phys. A}, 6:161--169, 1973.

\bibitem{Synge:1960ueh}
J.~L. Synge, editor.
\newblock {\em {Relativity: The General theory}}.
\newblock 1960.

\bibitem{Christensen:1979iy}
S.~M. Christensen and M.~J. Duff.
\newblock {Quantizing Gravity with a Cosmological Constant}.
\newblock {\em Nucl. Phys. B}, 170:480--506, 1980.

\bibitem{Prinz:2020nru}
David Prinz.
\newblock {Gravity-Matter Feynman Rules for any Valence}.
\newblock {\em Class. Quant. Grav.}, 38(21):215003, 2021.

\bibitem{Sannan:1986tz}
Sigurd Sannan.
\newblock {Gravity as the Limit of the Type {II} Superstring Theory}.
\newblock {\em Phys. Rev. D}, 34:1749, 1986.

\bibitem{DeWitt:1967uc}
Bryce~S. DeWitt.
\newblock {Quantum Theory of Gravity. 3. Applications of the Covariant Theory}.
\newblock {\em Phys. Rev.}, 162:1239--1256, 1967.
\newblock [,307(1967)].

\bibitem{Mertig:1990an}
R.~Mertig, M.~Bohm, and Ansgar Denner.
\newblock {FEYN CALC: Computer algebraic calculation of Feynman amplitudes}.
\newblock {\em Comput. Phys. Commun.}, 64:345--359, 1991.

\bibitem{Bilenky:1982tw}
Samoil~M. Bilenky.
\newblock {\em {Introduction to the Physics of Electroweak Interactions}}.
\newblock 1982.

\bibitem{Weinberg:1995mt}
Steven Weinberg.
\newblock {\em {The Quantum theory of fields. Vol. 1: Foundations}}.
\newblock Cambridge University Press, 6 2005.

\bibitem{Peskin:1995ev}
Michael~E. Peskin and Daniel~V. Schroeder.
\newblock {\em {An Introduction to quantum field theory}}.
\newblock Addison-Wesley, Reading, USA, 1995.

\bibitem{latosh2023}
Boris Latosh.
\newblock Operators present in one-loop scalar tensor gravity studied in
  arxiv:2304.08812 [gr-qc].
\newblock \url{https://doi.org/10.17632/zyn47cnsz3.1}, 2023.
\newblock Mendeley Data, V1.

\bibitem{Weinberg:1965nx}
Steven Weinberg.
\newblock {Infrared photons and gravitons}.
\newblock {\em Phys. Rev.}, 140:B516--B524, 1965.

\bibitem{Burgess:2003jk}
C.~P. Burgess.
\newblock {Quantum gravity in everyday life: General relativity as an effective
  field theory}.
\newblock {\em Living Rev. Rel.}, 7:5--56, 2004.

\bibitem{Bern:2019crd}
Zvi Bern, Clifford Cheung, Radu Roiban, Chia-Hsien Shen, Mikhail~P. Solon, and
  Mao Zeng.
\newblock {Black Hole Binary Dynamics from the Double Copy and Effective
  Theory}.
\newblock {\em JHEP}, 10:206, 2019.

\bibitem{Bern:2020buy}
Zvi Bern, Andres Luna, Radu Roiban, Chia-Hsien Shen, and Mao Zeng.
\newblock {Spinning black hole binary dynamics, scattering amplitudes, and
  effective field theory}.
\newblock {\em Phys. Rev. D}, 104(6):065014, 2021.

\bibitem{Chung:2018kqs}
Ming-Zhi Chung, Yu-Tin Huang, Jung-Wook Kim, and Sangmin Lee.
\newblock {The simplest massive S-matrix: from minimal coupling to Black
  Holes}.
\newblock {\em JHEP}, 04:156, 2019.

\bibitem{EventHorizonTelescope:2019dse}
Kazunori Akiyama et~al.
\newblock {First M87 Event Horizon Telescope Results. I. The Shadow of the
  Supermassive Black Hole}.
\newblock {\em Astrophys. J. Lett.}, 875:L1, 2019.

\bibitem{Bai:2016ivl}
Dong Bai and Yue Huang.
\newblock {More on the Bending of Light in Quantum Gravity}.
\newblock {\em Phys. Rev. D}, 95(6):064045, 2017.

\bibitem{Chi:2019owc}
Huan-Hang Chi.
\newblock {Graviton Bending in Quantum Gravity from One-Loop Amplitudes}.
\newblock {\em Phys. Rev. D}, 99(12):126008, 2019.

\end{thebibliography}

\end{document}